\documentclass{iopjournal}
\usepackage{graphicx}%
\usepackage[english]{babel}
\usepackage[autostyle, english = american]{csquotes}
\MakeOuterQuote{"}
\usepackage{multirow}%
\usepackage{cite}
\usepackage{amsmath,amssymb,amsfonts}%
\usepackage{amsthm}
\newtheorem{theorem}{Theorem}
\usepackage{enumitem}
\usepackage{parskip}
\usepackage{xcolor}
\usepackage[most]{tcolorbox}
\usepackage{ragged2e}
\usepackage{subcaption}
\renewcommand{\arraystretch}{1.4}
\justifying
\begin{document}

\articletype{Review} 

\title{Pulse Shaping for Superconducting Qubits}

\author{Animesh Patra$^1$, Ankur Raina$^1$\orcid{0000-0002-9022-3595}}
\vspace{-0.3cm}

\affil{$^1$Department of EECS, IISER Bhopal, Bhopal, India}
\vspace{-0.4cm}

\email{animeshpatra2059@gmail.com, ankur@iiserb.ac.in}
\vspace{-0.3cm}


\begin{abstract}
\justify{High-fidelity control of superconducting qubits requires carefully shaped microwave pulses to avoid several different kinds of error at once. 
This article is a pedagogical bridging text aimed at upper-level undergraduate and early graduate students who have completed an introductory quantum mechanics course and a first course in quantum computing or quantum information, but who have not yet encountered the physical implementation of qubit gates. 
We integrate physical intuition for pulse design, analytical gate-level descriptions, and practical hardware considerations into a single, derivation-driven narrative, with explicit learning objectives.
We begin with simple pulse envelopes and their spectral properties, showing how finite bandwidth produces leakage outside the computational subspace. 
This motivates the derivative removal by adiabatic gate (DRAG) technique, which we derive explicitly using the Magnus expansion, obtaining a clear, order-by-order account of which physical error channel appears at which order and why DRAG's cancellation is necessarily incomplete.
We discuss the practical hardware realities of control pulse generation, focusing on arbitrary waveform generators (AWG), local oscillators (LO), and IQ mixing. 
Finally, we extend the discussion to two-qubit operation via the cross-resonance gate, and interpret how driving the control qubit at the target qubit's transition frequency necessarily produces several unwanted interaction terms alongside the desired one, and how successive generations of pulse-engineering strategies have been designed to suppress them.} 
\end{abstract}

\section{Introduction}\label{sec1}
Quantum computers exploit the principles of quantum mechanics to perform computations. The expectation is that quantum computers will solve a certain class of intractable problems that classical devices cannot \cite{shor1999polynomial,arute2019quantum,bravyi2018quantum}. Central to their operation is the implementation of quantum gates at the hardware level \cite{song2025realization,spiteri2018quantum,haddadfarshi2016high,zhou2025high}. In practice, quantum gates are implemented through carefully designed control pulses that drive the dynamics of the underlying quantum system \cite{lloyd1993potentially,nakamura1999coherent,yu2002coherent}. Achieving high-fidelity gate operations is essential: even small deviations from ideal unitary evolution accumulate with circuit depth, ultimately degrading computational accuracy and limiting any potential quantum advantage.

In a college physics course setting, most introductory treatments of quantum computing emphasize abstract gate design and algorithmic structure, with comparatively little attention given to the physical realization of quantum gates. A student then trying to enter this line of research is faced with a series of questions: How does one actually implement a gate in practise? If one knows theoretically how to manipulate two-level systems, why is achieving high fidelity a hard task? What are the “moving parts” of the implementation, and how do they affect the outcome? How different are two-qubit gates from single-qubit gates in implementation?

The answers to the above questions lie in part in pulse shaping and sequencing. They play a fundamental role, effectively allowing one to engineer the system Hamiltonian by selectively enhancing desired terms while suppressing unwanted contributions. This allows the implementation of high-fidelity gates.

There exists a plethora of platforms to realize quantum computing \cite{kjaergaard2020superconducting,loss1998quantum,kane1998silicon,cirac1995quantum,childress2013diamond,nayak2008non,psaroudaki2023skyrmion} and even though the overarching principles governing pulse shaping remain largely uniform, the specific qubit platform of operation dictates the exact implementation. The superconducting platform has been one of the promising candidates in terms of scalability. Hence, in this article, we choose to focus on the superconducting qubit platform. This choice leads to microwave pulses and gate sets native to the superconducting platform. 

\begin{figure}[h]
\centering
\includegraphics[scale=0.6]{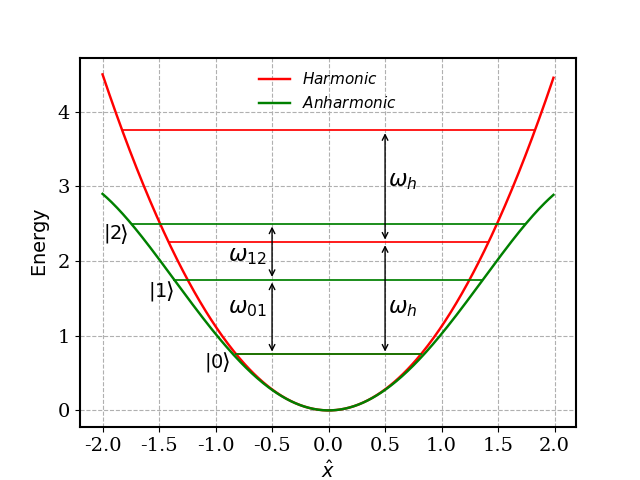}
\caption{\justifying  The potential energy profile and eigenenergies for the harmonic oscillator (solid red) and the anharmonic oscillator (solid green) in $\hbar=1$ units. The energy separation (equivalently, the transition frequency $\omega_{h}$) between the eigenstates of the harmonic oscillator is equal. For the aim of building a transmon qubit, an anharmonic oscillator is considered. Each energy separation (equivalently, the transition frequencies $\omega_{01}, \omega_{12}$ and so on) is different.
Differing energy separation allows to form a computational subspace from $|0\rangle$ and $|1\rangle$. 
While the rest of the states, like the $|2\rangle$ state, form the leakage subspace. } \label{HarmonicAnharmonic}
\end{figure}

A transmon qubit, widely used in this platform, is inherently a weakly anharmonic multi-level system \cite{superconducting_review}. A harmonic oscillator with its equally spaced multiple levels makes for a bad qubit candidate, as any resonant excitation always carries a large probability of leaking outside the qubit subspace and into non-computational states due to the equal transition frequency across multiple states as illustrated in Fig.\ref{HarmonicAnharmonic}. An anharmonic oscillator is then the natural choice to constrain the computational subspace in which the qubit exists from non-computational leakage states. However, the chance of leakage persists, and the transmon qubit remains far from an ideal two-level system. Consequently, pulse design must not only implement the desired qubit rotations but also suppress leakage into non-computational states. These make pulse shaping a nontrivial task, especially in the fast-gate regime.

Analytical pulse design offers the benefits of straightforward hardware implementation by reducing optimization overhead. However, obtaining the highest-fidelity pulse-shape solutions is often highly nontrivial and requires numerical techniques such as Gradient Ascent Pulse Engineering (GRAPE) \cite{khaneja2005optimal}. Nonetheless, studying analytical pulse design remains valuable for gaining insight into the physical system in question. To address this, we adopt the Magnus expansion \cite{blanes2010pedagogical, blanes2009magnus} as a systematic framework for analyzing driven quantum dynamics. Unlike standard time-dependent perturbation theory, the Magnus expansion provides a controlled, order-by-order description of the effective evolution operator, explicitly identifying the error terms at each order.
This insight enables the design of pulses that cancel specific unwanted contributions.
Using a three-level model that captures the minimal system with computational states and a single leakage state, we discuss the derivative removal by adiabatic gate (DRAG) technique \cite{motzoi2009simple} to mitigate leakage errors in quantum gates in the transmon qubit setup.

We examine some of the key hardware components involved in pulse shaping and how systematic errors and miscalibration distort the pulses, thereby reducing the fidelity of the quantum gate operation. We then extend the discussion to two-qubit operations. In contrast to the single-qubit case, simple pulse schemes are insufficient due to the presence of always-on interactions and additional error channels arising from a fixed, capacitively coupled transmon qubit architecture. We consider the cross-resonance (CR) gate, part of the native gate set for the fixed, capacitively coupled transmon architecture \cite{rigetti2010fully,chow2011simple}, which has emerged as a leading approach for scalable superconducting quantum processors.

Existing reviews serve complementary purposes but are aimed at audiences different from the one considered here. Comprehensive reviews such as those by Krantz et al.\cite{superconducting_review} provide an excellent overview of superconducting quantum computing, spanning qubit design, coherence, control, calibration, and readout. Their breadth, however, necessarily limits the depth with which pulse shaping itself can be developed as a teaching topic. Other reviews, such as those by Huang et al.\cite{huang2020superconducting}, emphasize superconducting-qubit architectures and readout mechanisms rather than the principles of pulse engineering. At the other end of the spectrum lie the original research papers that introduced pulse-shaping techniques, including the pioneering works of Motzoi et al.\cite{motzoi2009simple}. These articles present mathematically rigorous derivations and optimization strategies intended primarily for researchers already familiar with driven quantum systems, rotating frames, microwave control hardware, and effective Hamiltonians.

As a result, students are often introduced to pulse-shaping methods as a collection of successful engineering techniques without first developing a coherent understanding of why those techniques emerge from the underlying physics. They may learn that Gaussian pulses reduce spectral leakage, that DRAG suppresses leakage errors, or that echoed cross-resonance sequences improve two-qubit gates, yet the common physical principles linking these developments often remain implicit.

This article addresses that pedagogical gap by providing a conceptual bridge between the abstract quantum gates encountered in introductory quantum computing courses and their physical implementation in superconducting qubits. Rather than presenting individual pulse-engineering techniques as isolated protocols, we organize the discussion around the common framework of Hamiltonian engineering and systematic error suppression. In Sec. \ref{sec3}, we introduce the basic principles of pulse shaping, beginning with Rabi oscillations in a two-level system \cite{rabi1937space}. We progressively introduce finite-bandwidth effects, leakage in weakly anharmonic systems, and proceeding to a three-level analysis using the Magnus expansion, which motivates the DRAG protocol. In Sec. \ref{sec4}, we discuss the hardware components involved in pulse generation, along with common sources of control errors arising from imperfect signal synthesis. Finally, in Sec. \ref{sec5}, we examine two-qubit gates, focusing on the pulse-engineering strategies required for high-fidelity cross-resonance operations. Throughout, each successive technique is motivated as a natural response to the error channels identified in the preceding analysis.

\subsection{Learning objectives}
Having worked through this article, a reader
should be able to:
\begin{itemize}
\item[\textbf{L1}] Use the Magnus expansion to determine, order by order, which error channels arise in a realistic multi-level qubit setup, and explain why this is preferable to a naive truncation of time-dependent perturbation theory.
\item[\textbf{L2}] Derive the DRAG quadrature condition (Eq.~\ref{DRAG Substitution}) from the requirement that the first-order leakage term vanishes, and explain physically why the DRAG correction cannot remove leakage at all orders.
\item [\textbf{L3}] Learn the utility of each hardware component involved in the implementation of a pulse and relate the hardware error channels to Pauli terms.
\item[\textbf{L4}] Give an intuitive explanation of the effective cross-resonance Hamiltonian of Eq.~\ref{CR Hamiltonian}, why driving the control qubit at a target qubit's transition frequency produces a $Z\otimes X$ interaction accompanied by several unwanted terms, and describe at least two distinct pulse-engineering strategies used to
suppress them.
\end{itemize}

\subsection{Prerequisites}

Beyond the background above, the reader should be comfortable with: (i) the Pauli operator formalism of a two-level system; (ii) the Schr\"odinger equation and the construction of unitary evolution operators, including changes of frame via interaction-picture transformations; (iii) elementary Fourier analysis; and (iv) the standard circuit model of quantum computation, including single-qubit gates and the CNOT gate, at the level of \cite{nielsen2010quantum}. 
No prior exposure to the Magnus expansion is assumed. 
We introduce it in Sec \ref{subsec22}. We do not assume any prior exposure to microwave engineering or laboratory control electronics.  \\
\vspace{0.25cm}


\section{Physics of Pulse Shaping}\label{sec3}
\vspace{0.25cm}
At its core, pulse shaping concerns the controlled dynamics of driven quantum systems, addressing how the temporal profile of an applied driving field influences quantum evolution. Different pulse envelopes—such as square, Gaussian, triangular, or modulated pulses—possess distinct spectral characteristics, leading to different transition probabilities, leakage mechanisms, and coherence properties. This section develops the physical principles underlying pulse shaping that later motivate engineering considerations and hardware-aware refinements.

We lay out the following objectives for the reader:
\begin{enumerate}[label=(\roman*)]
    \item Explain using Fourier analysis the trade-off between pulse duration and spectral bandwidth, and the influence on unwanted transitions.
    \item Describe how coherent control is achieved in an ideal two-level system to implement single-qubit gate operations.
    \item Use the Magnus expansion to identify dominant error channels in a systematic, order-by-order manner.
    \item Analyzing the three-level model as the minimal entry point to gauge into the multi-level transmon setup.
    \item Derive the DRAG condition and interpret it as a Hamiltonian-engineering strategy for suppressing the dominant leakage channel.
\end{enumerate}

\subsection{Rabi Oscillations in Two-Level Systems}\label{subsec31}

An ideal qubit is a two-level system. 
Hence, it is appropriate to use the Pauli matrices $\sigma_{x}, \sigma_{y}$ and $\sigma_{z}$ to describe the qubit's dynamics. For quantum computation, one needs to manipulate the qubit's state. 
Hence, it is in our interest to investigate what happens to the qubit when it is subjected to a controlled external influence, specifically driving in the transverse direction.
Therefore, we write the qubit Hamiltonian as in Eq. \ref {rabi two level}. 
The Hamiltonian describes a driven two-level setup. 
The first term is the static part, describing the qubit's separation between its two states with qubit transition frequency $\omega_{q}$. The second term is a time-dependent term with the driving (or pulse-carrier) frequency $\omega_{d}$ and the drive phase $\alpha$. 
Without loss of generality, we take $\sigma_{y}$ in the second term to represent the transverse drive. Importantly to us, the pulse amplitude and shape information is encoded in the function $A(t)$, 
\begin{equation}\label{rabi two level}
	H = \frac{-\hbar\omega_{q}}{2}\sigma_{z} + \hbar A(t)\sin(\omega_{d}t+\alpha)\sigma_{y}.
\end{equation}

\begin{tcolorbox}[colback=gray!5!white, colframe=gray!75!black, title= Lab Frame to Rotating Frame]

Instead of working in the lab frame, we can transform into the rotating frame. 
A valid confusion among students is why transform to another frame if the underlying physics remains the same? The reason is convenience and simplicity. 
Imagine describing your state from outer space; your motion would include the Earth's daily rotation and yearly revolution, making even standing still appear rather complicated. 
In contrast, describing your position from the Earth's frame is much simpler because the dominant motion has already been accounted for. 
Similarly, by transforming to the oscillatory drive’s frame or the rotating frame, we aim to isolate the slowly-varying dynamics that govern the qubit's evolution. 
The transformation is exact, and the physics remains the same, but the Hamiltonian becomes easier to interpret. 
In the rotating frame, the Hamiltonian generally contains both slowly varying and rapidly oscillating terms; however, in most cases, these can be neglected by applying the rotating wave approximation.
\end{tcolorbox}
\vspace{0.25cm}
We define a unitary operator $V(t)$ that transforms from the lab frame to the rotating frame, $V(t)|\Psi(t)\rangle=|\Psi(t)\rangle_{\mathrm{rot}}$. 
Consider the Schrödinger equation,
\begin{equation}
\begin{split}
        \iota\hbar\frac{\partial}{\partial t}|\Psi(t)\rangle &= H(t)|\Psi(t)\rangle,\\
        \iota\hbar\frac{\partial}{\partial t}\{V^{-1}(t)|\Psi(t)\rangle_{\mathrm{rot}}\} &= H(t)V^{-1}(t)|\Psi(t)\rangle_{\mathrm{rot}}\,.
\end{split}\label{SE 1}
\end{equation}

The action of $V(t)$ from the left on both sides of the equation leads to,
\begin{equation}
    \begin{split}
       \iota\hbar\frac{\partial}{\partial t}|\Psi(t)\rangle_{\mathrm{rot}} &= \{V(t)H(t)V^{-1}(t)-\iota\hbar V(t)\frac{\partial}{\partial t}V^{-1}(t)\}|\Psi(t)\rangle_{\mathrm{rot}}\,\\
       &= H_{\mathrm{rot}}|\Psi(t)\rangle_{\mathrm{rot}}\, .
    \end{split}\label{SE 2}
\end{equation}

Now, let us consider the qubit Hamiltonian (Eq.\ref{rabi two level}) and transform into the rotating frame of the drive. The rotating frame Hamiltonian with $V(t) = \exp(\frac{-\iota \omega_{d}t}{2}\sigma_{z})$ is
\begin{equation}
    \begin{split}
        H_{\mathrm{rot}} &=-\frac{\hbar\delta}{2}\sigma_{z}+\frac{\hbar A(t)}{2}\{\sigma_{y}\sin(\alpha)-\sigma_{x}\cos(\alpha)\}\\
        &+\frac{\hbar A(t)}{2}\{\sigma_{y}\sin(2\omega_{d}t+\alpha)+\sigma_{x}\cos(2\omega_{d}t+\alpha)\}.
    \end{split}
\end{equation}\label{calc 1}

Here $\delta=\omega_{q}-\omega_{d}$ is the detuning of the drive or equivalently the mismatch between the qubit transition frequency and the frequency of the drive. Now, we invoke an important approximation, the \textit{rotating wave approximation (RWA)}, allowing us to drop the fast oscillatory term. Further, without loss of generality, we choose $\alpha=\pi$. So the RWA Hamiltonian becomes,
\begin{equation}\label{rwa ham}
    H_{\text{RWA}} = -\frac{\hbar\delta}{2}\sigma_{z}+\frac{\hbar A(t)}{2}\sigma_{x}\, .
\end{equation}

\begin{tcolorbox}[breakable,colback=gray!5!white, colframe=gray!75!black, title= The Rotating Wave Approximation (RWA)]

To any student, it may seem confusing or even feel like cheating to downright discard the fast-rotating term from the rotating frame Hamiltonian. A natural question is, “Why is the approximation valid?” The answer lies in the separation of timescales between the drive envelope and the carrier oscillation. The term $\Omega(t)\mathrm{sin}(\omega_{d}+\omega_{t}\;t)$ can be ignored if,
\begin{equation}\label{relative_strength}
    \frac{\Omega_{max}}{\omega_{d}+\omega_{t}}<<1.
\end{equation}
This means that the pulse envelope, whose characteristic time evolution $\mathrm{T}$ is of the order $\frac{1}{\Omega_{max}}$, varies only slightly during one cycle of the rapidly oscillating term. Consequently, the envelope may be regarded as approximately constant over each fast oscillation. The pulse contains multiple cycles of the fast oscillatory terms whose contributions average out to zero in this regime. This can be seen as follows: consider the cycle duration of the fast oscillations as $t_{f}=\frac{1}{\omega_{d}+\omega_{t}}$ then,
\begin{equation}
    \int_{0}^{T}\Omega(t)e^{\iota (\omega_{d}+\omega_{t})t}dt\approx \sum_{n}\Omega(nt_{f})\int_{nt_{f}}^{(n+1)t_{f}}e^{\iota (\omega_{d}+\omega_{t})t}=0
\end{equation}
We, however, run into a problem when the relative strength is of the order of unity. Then, the pulse changes appreciably over the cycle of fast oscillations, and the pulse cannot be approximated as piecewise constant during the fast oscillatory cycle. Such a scenario arises for ultra-fast gates. 
Shorter gate times mean stronger pulses are required, which means that in such a regime, the validity of the approximation becomes dubious.
The real approximation, then, is not the fact that the fast oscillatory terms average out to zero, but rather it is an approximation about the physical regime in which the system is operated.
\end{tcolorbox}
\vspace{0.25cm}
To study the dynamics, we first need to determine the unitary evolution operator corresponding to the Hamiltonian in Eq. \ref{rwa ham}. 
Even though the Hamiltonian in Eq. \ref{rwa ham} is vastly simplified with no fast oscillatory terms, the time dependence from the function $A(t)$ still makes it difficult to derive an exact expression for any generic pulse shape $A(t)$. 
Hence, we resort to approximations. We introduce the Magnus expansion, which helps in obtaining an effective evolution operator.

\subsection{Magnus Expansion}\label{subsec22}

The Magnus expansion is not just an alternative to time-dependent perturbation theory. Rather, it is an improvement in the sense that truncating the Magunus expansion at some finite order always preserves the unitarity. Most importantly, in the context of pulse shaping, the commutator terms that arise at each order give us information on what kinds of errors arise exactly at that order. 
This helps build some intuition about which errors might be "more erroneous" than the others. This introduction of the Magnus expansion is by no means complete, and the reader should check out further texts to appreciate its utility in different settings \cite{blanes2009magnus}.

Mathematically, the expansion assumes that a true exponential solution exists for the evolution operator of a time-dependent Hamiltonian,
\begin{equation}
    \begin{split}
        U(t,t_{0}) = \exp(\Omega(t,t_{0})),\qquad\Omega(t_{0},t_{0})=0\,,\qquad\qquad\qquad
    \end{split}\label{magnus intro}
\end{equation}
and the abstract operator $\Omega$ in the exponent can itself be written as a series expansion,
\begin{equation}
    \begin{split}
        \Omega(t,t_{0}) = \sum_{i}^{\infty}\Omega_{i}(t,t_{0}). 
    \end{split}\label{magnus intro 2}
\end{equation}
One then proceeds to find and solve the series expansion for $\Omega(t,t_{0})$.

We do not go into the derivations, but we provide forms for the first few orders, and we shall discuss their meaning in the context of qubit pulse shaping. 
Though solving the integrals in the expression provides us with exact coefficients for a particular order, we are more interested in the structure of the integrand and the operators arising
from the commutators.
We set $-\frac{\iota}{\hbar}H=\tilde{H}$,
\begin{equation}\label{magnus formulae}
    \begin{split}
        \Omega_{1}(t,t_{0})&=\int_{t_{0}}^{t}dt'\tilde{H}(t'),\\
        \Omega_{2}(t,t_{0})&=\frac{1}{2}\int_{t_{0}}^{t}dt'\int_{t_{0}}^{t'}dt''[\tilde{H}(t'),\tilde{H}(t'')],\\
        \Omega_{3}(t,t_{0})&=\frac{1}{6}\int_{t_{0}}^{t}dt'\int_{t_{0}}^{t'}dt''\int_{t_{0}}^{t''}dt'''[\tilde{H}(t'),[\tilde{H}(t''),\tilde{H}(t'')]]+[\tilde{H}(t'''),[\tilde{H}(t''),\tilde{H}(t')]].
    \end{split}
\end{equation}

With the Magnus expansion at our disposal, we now proceed to find a general expression for the first and second order expansion terms for the Hamiltonian in Eq. \ref{rwa ham}.
The first-order Magnus term reads,
\begin{equation}\label{general first order}
    \Omega_{1}(T) = -\frac{\iota}{2}\Big(\delta T\sigma_{z}+\int_{0}^{T}dt\;A(t)\sigma_{x}\Big).
\end{equation}
Similarly, the second-order Magnus term reads
\begin{equation}\label{general second order}
    \Omega_{2}(T)=\frac{\iota}{8}\delta\sigma_{y}\Big(\int_{0}^{T}dt\int_{0}^{t}dt'\big\{A(t')-A(t)\big\}\Big).
\end{equation}

Even before we go into specific examples, we can make two important observations. For two level system,
\begin{enumerate}[label=(\roman*)]
    \item The first-order term only involves the area under the pulse shape. Two differing shapes with the same area will generate the same first-order term in Eq. \ref{general first order}. 
    \item Non-trivial pulse-shaping effects only enter at the second order. Non-trivial pulse-shaping effects only enter at the second order. For a generic pulse shape, $A(t')\ne A(t)$ at some arbitrary time $t'\ne t$. Hence, the integrand in Eq. \ref{general second order} might be non-zero and contribute to the effective evolution operator.
    \item The explicit dependence of the second-order term on detuning $\delta$ in Eq. \ref{general second order} implies that at resonance ($\delta=0$), all higher-order terms evaluate to zero. Hence, pulse shaping has no explicit effect for two-level systems at resonance.
\end{enumerate}  

\begin{tcolorbox}[colback=gray!5!white, colframe=gray!75!black, title= More levels is different]
While the interpretation of the Magnus expansion equations we get for the two-level system is correct, they are by no means a pattern for a multi-level system. Textbooks generally only discuss two-level systems, and hence, readers often incorrectly assume these inferences as the rule of thumb. In reality, just by introducing a third level,
\begin{enumerate}[label=(\roman*)]
    \item New explicit detuning- or anharmonicity-dependent terms start appearing at first order itself. These may be the leakage channels that we want to mitigate.
    \item Pulse shape becomes physically significant even on resonance. Although the carrier is resonant with the computational transition, the pulse envelope possesses a finite bandwidth. Spectral components of the pulse can therefore excite nearby transitions, producing leakage and phase errors. 
    \item Higher-order terms remain important. Resonance removes the static detuning of the intended transition, but it does not eliminate higher-order contributions arising from the multilevel structure.
\end{enumerate}
\end{tcolorbox}

\subsubsection{Example}

We now consider two pulse shapes to study Rabi oscillations and demonstrate the usage of Magnus expansion. The square pulse of duration $T$ is given by,
\begin{equation}\label{square pulse}
    A(t) = A_{0}\quad \text{for}\quad t\in[0,T],
\end{equation}
while the triangular pulse is given as,
\begin{equation}\label{triangular pulse}
    A(t) = \begin{cases}
    \frac{2A_{0}}{T}t&t\in[0,\frac{T}{2}]\\
    \frac{2A_{0}}{T}(T-t)&t\in[\frac{T}{2},T].
    \end{cases}
\end{equation}

\begin{figure}
\centering
\begin{subfigure}{0.5\textwidth}
  \centering
  \includegraphics[scale=0.48]{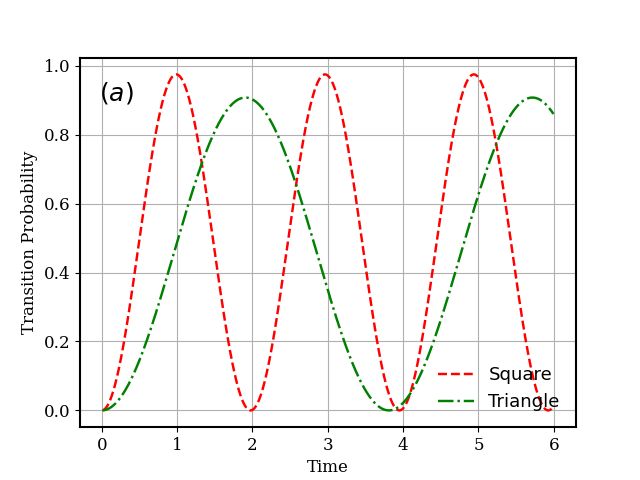}
\end{subfigure}%
\begin{subfigure}{0.5\textwidth}
  \centering
  \includegraphics[scale=0.48]{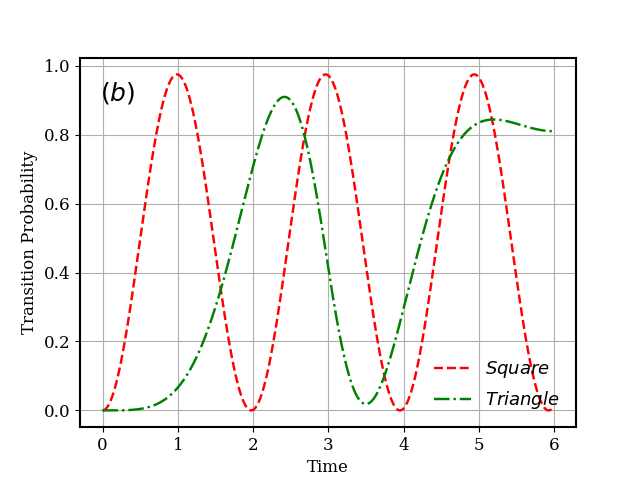}
\end{subfigure}
\caption{(a)The transition probability of the square and the triangular pulse from Magnus expansion truncated at second order for $\delta =0.5,\;A_{0}=\pi$. (b) Numerical simulation of exact dynamics shown by RWA Hamiltonian of Eq.\ref{rwa ham} for square and triangular pulse at $\delta =0.5,\;A_{0}=\pi$. The simulation is carried out using QuTiP in Python.}\label{ApproxExactComparision}
\end{figure}

For the square pulse, the second-order Magnus term (Eq. \ref{general second order}) is zero due to the pulse shape being a constant. 
The first order expression for the square pulse $\Omega_{1}^{\mathrm{sq}}(T)$ is,
\begin{equation}\label{Square Magnus}
    \Omega_{1}^{\mathrm{sq}}(T) =-\frac{\iota T}{2}\Big(\delta\sigma_{z}+A_{0}\sigma_{x}\Big). 
\end{equation}

So the evolution operator becomes $U(T) = \exp\big(\Omega_{1}^{\mathrm{sq}}(T)\big)=\exp(-\frac{\iota \omega_{\mathrm{sq}}T}{2}\;\vec{n}.\vec{\sigma})$. 

Here, $\omega_{\mathrm{sq}}=\sqrt{\delta^{2}+A_{0}^{2}}$ and $\vec{n}\equiv(\frac{A_{0}}{\omega_{\mathrm{sq}}},0,\frac{\delta}{\omega_{0}})$, where $\vec{n}$ is a three-dimensional unit vector and $$\vec{n}.\vec{\sigma}=n_{x}\sigma_{x}+n_{y}\sigma_{y}+n_{z}\sigma_{z}.$$
We are concerned with the manipulation of the qubit's state. 
So, it is natural to look at the probability of finding the state in the excited state if the qubit was initialized in the ground state.
So, let us look at the transition probability $P_{0\xrightarrow{}1}(t)$,
\begin{equation}
    \begin{split}\label{transition prob 1}
        P_{0\xrightarrow{}1}(T) &= |\langle1|U(T)|0\rangle|^{2}\\
        &=\frac{A_{0}^{2}}{A_{0}^{2}+\delta^{2}}\sin^{2}\Big(\frac{\omega_{\mathrm{sq}}T}{2}\Big).
    \end{split}
\end{equation}

Here we used the identity $\exp(+\iota\theta\vec{n}.\vec{\sigma})=\cos(\theta) I + \iota\sin{(\theta)}\vec{n}.\vec{\sigma}$. One can see that we have a non-zero probability of transition from Eq. \ref{transition prob 1}, and that it is periodic. 
This is \textit{Rabi-oscillations}. 
If one lets the pulse operate for time $T=\frac{\pi}{\omega_{\mathrm{sq}}}$, the pulse attains its maximum transition, and this pulse is called a $\pi$-pulse. More generally, a $\pi$-pulse is any pulse whose integrated action produces a $\pi$ rotation of the qubit state, irrespective of its temporal shape. We note that in the special case of resonant driving, one can achieve a complete transition.

For the triangular pulse case, we consider only up to the second order of Magnus expansion. These are given by,
\begin{equation}\label{triangle first order}
    \Omega_{1}^{\mathrm{tr}}(T) = -\frac{\iota T}{2}\Big(\delta\sigma_{z}+\frac{A_{0}}{2}\sigma_{x}\Big),
\end{equation}
\begin{equation}\label{triangle second order}
    \Omega_{2}^{\mathrm{tr}}(T) = \Big(\frac{\iota \delta}{8}\Big)\Big(\frac{A_{0}}{12}\Big)\Big(\frac{T}{2}\Big)^{2}\sigma_{y}.
\end{equation}
The corresponding unitary then becomes $U(T) = \exp\big(\Omega_{1}^{\mathrm{tr}}(T)+\Omega_{2}^{\mathrm{tr}}(T)\big)$. After some algebraic manipulations and bringing it to the $\exp(+\iota\theta\vec{n}.\vec{\sigma})$ form, the transition probability for the triangular pulse case becomes
\begin{equation}
    P_{0\xrightarrow{}1}^{\mathrm{tr}}(t) = \Big(1-\frac{\big(\delta T\big)^{2}}{\big(\delta T\big)^{2}+\big(\frac{A_{0}T}{2}\big)^{2}+\big(\frac{\delta A_{0}T^{2}}{192}\big)^{2}}\Big)\sin^{2}\Big(\sqrt{\big(\delta T\big)^{2}+\big(\frac{A_{0}T}{2}\big)^{2}+\big(\frac{\delta A_{0}T^{2}}{192}\big)^{2}}\Big).
\end{equation}

The evolution operator generated from truncated Magnus expansion at the second order approximates the exact dynamics closely in the small $\delta$ regime as can be seen in Fig. \ref{ApproxExactComparision}. Away from the resonant driving point, depending on the pulse shape and driving parameter values, certain driving protocols are better suited to induce transitions across the two-level system. 
In this example, the square pulse appears to induce transitions to the excited state with higher probability, given that we started from the ground state.
However, the triangular pulse's transition probability profile is smoother, providing a better safeguard against random perturbations that can arise in real-life experiments.

With this example, the use of the Magnus expansion might be clearer, and the role of the pulse shape in the dynamics more appreciated. 
Things become more involved and interesting when leakage states are also present in the system; that is, off-resonant transitions to out-of-computational subspace states become possible. 
Then, the pulse's spectral properties also come into play, and the square pulse, which was "better" in our ideal two-level system example, may become "worse" due to its broad frequency spectrum.
Thus, enabling more off-resonant transitions than, say, the triangular pulse, which is smoother than the square pulse.

Even in this simple description, it is important to note that for a pulse period $T$, any amplitude errors from the hardware affecting $A(t)$ lead to a slightly different state than expected.
Such small deviations can accumulate over time as more gates are applied.
Hence, it is important to study errors originating from hardware itself.
We will see this in Sec.  \ref{sec4}.

\subsection{Three-Level Model}\label{subsec33}

We now consider a three-level system. The motivation to consider this is that one does not have access to an ideal two-level system in the transmon qubit setup. 
The qubit is not perfectly isolated, so nearby energy levels get unintentionally excited.
In most cases, experimentalists use control methods to create an effectively two-level system from a many-level system. 
To study the deviations from the ideal two-level system, the three-level system is interpreted as a two-level system with an additional leakage state.

Consider the Hamiltonian $H_{0}$, which represents a system with unequal level spacing (in line with the three lowest levels of the anharmonic oscillator in the laboratory setup),
\begin{equation}
H_{0}=\hbar \begin{pmatrix}
            0& \;0&0\\
            0&\;\omega_{1}&0\\
            0&\;0&\;\omega_{2}
            \end{pmatrix}.\label{3lvl static H}
\end{equation}
\begin{figure}[h]
\centering
\includegraphics[scale=0.6]{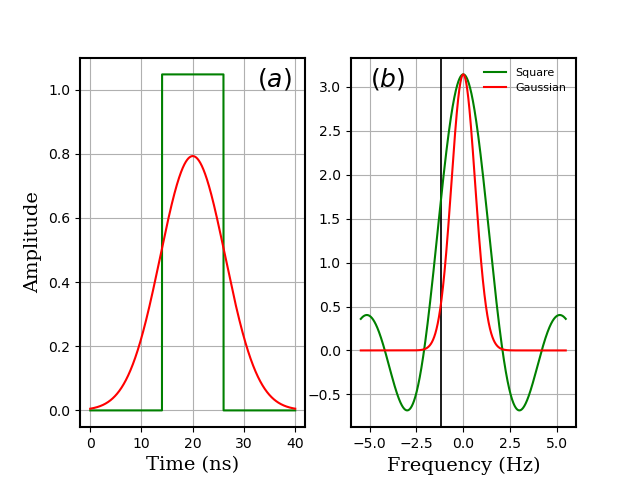}
\caption{\justifying (a) The square pulse (green line) and the Gaussian pulse (red line). The square pulse is analytically easier to study however, (b) it's baseband frequency spectrum has wider sidelobes than the Gaussian pulse. }\label{Pulse Shapes FFT}
\end{figure}
\begin{figure}[h]
\centering
\includegraphics[scale=0.65]{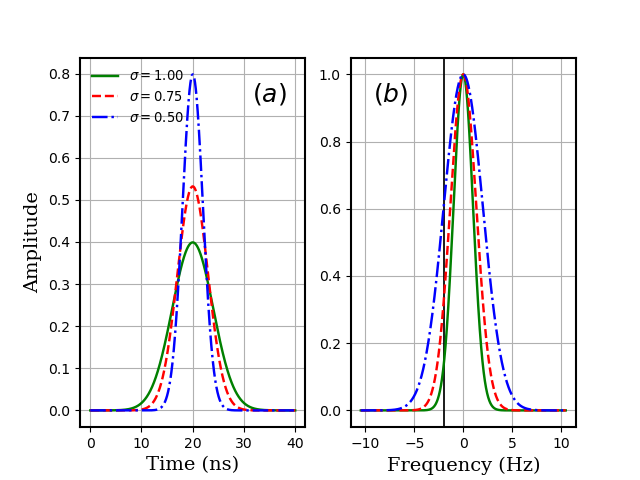}
\caption{\justifying (a) The solid-green, dashed-red, and dash-dotted blue lines are all Gaussian pulses with decreasing pulse duration. (b) Shorter pulses have a wider frequency spread. Broader frequency support increases the chances of overlap with the unwanted transition frequency. }\label{Time-Frequency Comparision}
\end{figure}
Even before we push ahead to the calculations, we can visually understand the problem at hand. Let us consider two envelope shapes: the square pulse and the Gaussian pulse envelope. Looking at the Fourier spectra, one can clearly see that the spectrum is not a sharp peak. 
Rather, it has a significant spread at neighboring frequencies. 
If the pulse shape has significant weight at the frequency corresponding to the transition between the qubit's excited state and the leakage state, as indicated by the black line in Fig. \ref{Pulse Shapes FFT}(b), it would naturally lead to a higher probability of leakage errors. The problem gets worse for a smaller anharmonicity $\Delta$, that is, the deviation from the equal energy level spacing for a pair of energy levels. There is only so much anharmonicity one can add to the system. We see that different pulses can have different weights at different frequencies, leading to some being better than others.

Further, we want our computations to be fast. Hence, our quantum gate operations should be fast too. However, shorter pulses also provide broader frequency support as seen in Fig.\ref{Time-Frequency Comparision}(b). Broader frequency support can lead to significantly more weight at the undesired transition frequency. These two observations show the importance of understanding pulse shaping.

Now we try to analyze the three-level system to pinpoint exactly which types of errors arise and in what order. The structure of the Magnus expansion terms may help us understand the pulse shaping strategy. Let us look at the driving Hamiltonian $H_{d}$ (Eq.\ref{3lvl drive}). Here, we have two controls on the driving through the in-phase component $I(t)$ and the quadrature component $Q(t)$. $\sigma_{k,k+1}^{+}$ and $\sigma_{k,k+1}^{-}$ promote transitions to neighbouring levels $k$ and $k+1$, while $\lambda_{k,k+1}$ is the coupling constants between the levels $k$ and $k+1$. Without loss of generality we can assume $\lambda_{0,1}=1$ and $\lambda_{1,2}=\lambda$, 
\begin{equation}
H_{d} = \hbar \Big(I(t)\cos{(\omega_{d} t)}+Q(t)\sin{(\omega_{d} t)}\Big)\sum_{k=0,1}\lambda_{k,k+1}(\sigma_{k,k+1}^{+}+\sigma_{k,k+1}^{-}).\label{3lvl drive}
\end{equation}

Now let us transform to the interaction picture, and then apply the RWA, thereby neglecting all fast-moving terms relative to the qubit's resonant transition ($\omega>\omega_{1}$). We choose $V = \exp{(+\frac{\iota}{\hbar}H_{0}t)}$,
\begin{equation}
V\sigma_{01}^{\pm}V^{-1}=e^{\pm\iota\omega_{1}t}\sigma_{01}^{\pm},\qquad
V\sigma_{12}^{\pm}V^{-1}=e^{\pm\iota(\omega_{2}-\omega_{1})t}\sigma_{12}^{\pm}\label{3lvl creation}
\end{equation}
If the system is driven resonantly with the qubit frequency, that is, $\omega_{d}=\omega_{1}$, then the leakage state detuning $\delta_{2}$ is equivalent to the anharmonicity in the system $\Delta = \omega_{2}-2\omega_{1}$. Defining $\sigma_{01}^{x}=\sigma_{01}^{+}+\sigma_{01}^{-}$ and $\sigma_{01}^{y}=\iota\sigma_{01}^{+}-\iota\sigma_{01}^{-}$. Consequently, we get the following,
\begin{equation}
\begin{split}
H_{RWA} &= H_{01} + H_{12},\\
H_{RWA} &= \frac{\hbar}{2}\Big(I(t)\sigma_{01}^{x}-Q(t)\sigma_{01}^{y}\Big)\\
    &+\frac{\lambda\hbar}{2}\Big[\Big(I(t)
    +\iota Q(t)\Big)e^{+\iota \Delta t}\sigma_{12}^{+}
    +\Big(I(t)
    -\iota Q(t)\Big)e^{-\iota \Delta t}\sigma_{12}^{-}\Big]. 
\end{split}\label{3 lvl RWA}
\end{equation}

\begin{figure}[h]
\centering
\includegraphics[scale=0.65]{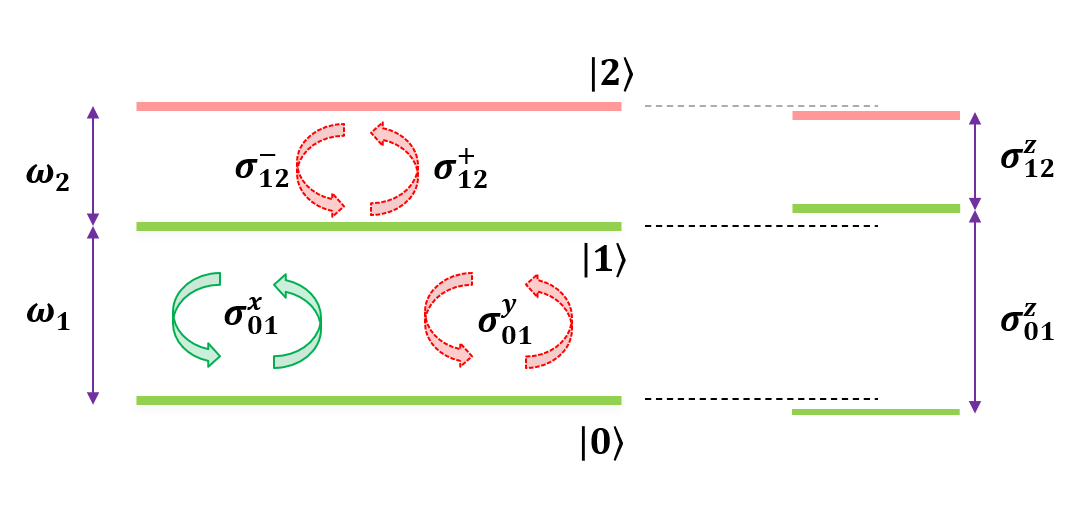}
\caption{\justifying A schematic of the three-level model. The first two levels form the computational subspace with a transition frequency of $\omega_{1}$. The $\sigma_{01}^{x}$ and $\sigma_{01}^{y}$ promote transition inside the computational subspace. However, the $\sigma_{01}^{y}$ contribution is the unwanted term arising in the first-order Magnus expansion. The DRAG protocol helps eliminate this and the coupling to the leakage state (arrows shown in red). 
AC Stark effect contributions emerge at the second order in the expansion ($\sigma_{01}^{z}$ and $\sigma_{12}^{z}$ terms).}\label{Three Level Schematic}
\end{figure}
$H_{\text{RWA}}$ is split into $H_{01}$ and $H_{12}$ in Eq.\ref{3 lvl RWA}.
The former is related to the computational subspace, and the latter is related to transitions between the excited and the leakage state.
Now we apply the Magnus expansion on our three-level $H_{RWA}$. The first order term is as follows,
\begin{equation}\label{1order magnus 1}
    \begin{split}
        \Omega_{1}(T) &= -\frac{\iota}{\hbar}\int_{0}^{T}dt\; H_{01}+H_{12}=A_{1} + A_{2},\\
        A_{1} &= -\frac{\iota}{2}\left( \int_{0}^{T} I(t)\;dt\;\right)\;\sigma_{01}^{x} + \frac{-\iota}{2}\left(\int_{0}^{T} Q(t)\;dt\;\right)\;\sigma_{01}^{y},\\
    \end{split}
\end{equation}
\begin{equation}\label{1order magnus 2}
    \begin{split}
        A_{2} &= -\frac{\iota\lambda}{2}\left( \int_{0}^{T} dt\;\Big(I(t)+\iota Q(t)\Big)e^{+\iota \Delta t}\;\sigma_{12}^{+} + h.c.\right)\\
        &= A_{+} + A_{-}.
    \end{split}
\end{equation}
Clearly, at the first order itself, there are imperfections.
There exists a $\sigma_{01}^{y}$ term in $A_{1}$ which contributes to undesirable rotations during the dynamical evolution, deviating from the desired $X$ gate. 
Further terms, such as $\sigma_{12}^{+}$ and $\sigma_{12}^{-}$ in $A_{2}$, promote the transition from the excited state of the computational subspace to the leakage state (Fig.\ref{Three Level Schematic}). 
This prompts the question: can one suppress or eliminate these?
We can, in fact, do that by analyzing the integrand's structure.
Let us set the quadrature $Q(t)$ as proportional to the derivative of the in-phase signal $I(t)$. To be exact, we set,
\begin{equation} 
    \begin{split}
        Q=-\frac{\dot{I}(t)}{\Delta},
    \end{split}\label{DRAG Substitution}
\end{equation}
and assume that $I(0)=I(T)=0$, that is, there is no driving effect after the pulse period. Let us look into the term $A_{+}$. Applying integration by-parts on the $Q(t)$ part in Eq.\ref{integration},

\begin{equation}
    \begin{split}
        A_{+} &= \frac{-\iota\lambda\sigma_{12}^{+}}{2}\left(\int_{0}^{T}I(t)e^{+\iota\Delta t}dt + \iota e^{+\iota\Delta t}\int Q(t) dt \;\Bigg|_{0}^{T} +\Delta\int_{0}^{T} dt\;e^{+\iota\Delta t}\int dt'\;Q(t')\right)\\
        & = \frac{-\iota\lambda\sigma_{12}^{+}}{2}\left(\int_{0}^{T}I(t)e^{+\iota\Delta t} - \iota e^{+\iota\Delta t}\int \frac{\dot I(t)}{\Delta} dt \;\Bigg|_{0}^{T} -\Delta\int_{0}^{T} dt\;e^{+\iota\Delta t}\int dt'\;\frac{\dot I(t')}{\Delta}\right)\\
        &= \frac{-\iota\lambda\sigma_{12}^{+}}{2}\left(\int_{0}^{T}I(t)e^{+\iota\Delta t} - \Big(\frac{\iota}{\Delta} e^{+\iota\Delta t}I(t) \;\Bigg|_{0}^{T}\Big) -\int_{0}^{T} dt\;I(t)e^{+\iota\Delta t}\right)\\
        &= \frac{+\iota\lambda\sigma_{12}^{+}}{2}\left( \frac{\iota}{\Delta} e^{+\iota\Delta T}I(T)-\frac{\iota}{\Delta}I(0)\right) = 0.
    \end{split}\label{integration}
\end{equation}
\begin{figure}
\centering
\begin{subfigure}{0.5\textwidth}
  \centering
  \includegraphics[scale=0.48]{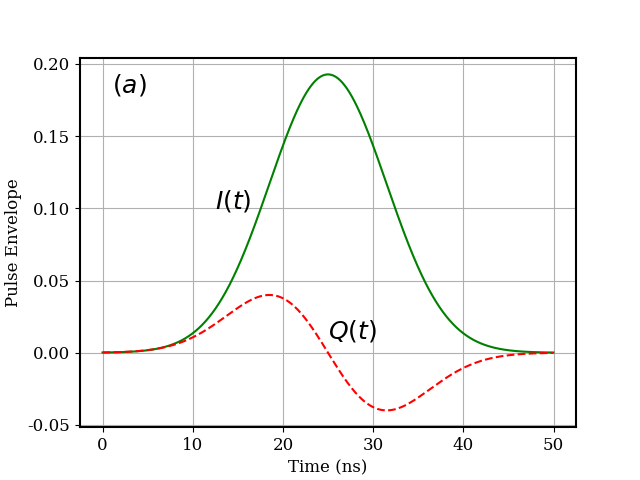}
\end{subfigure}%
\begin{subfigure}{0.5\textwidth}
  \centering
  \includegraphics[scale=0.48]{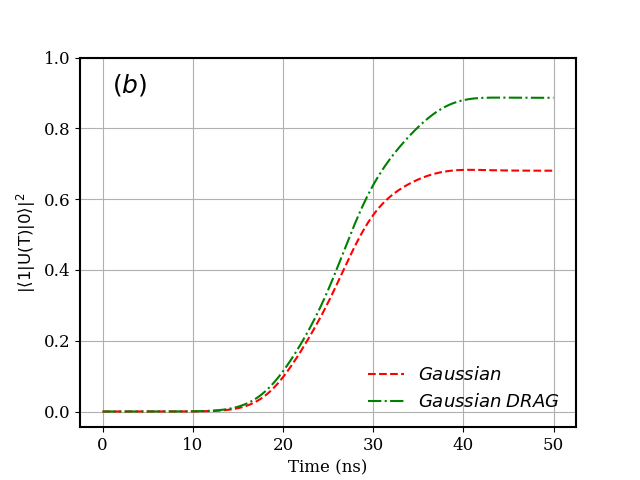}
\end{subfigure}
\caption{(a)A schematic illustrating the in-phase I(t) and the quadrature Q(t) component for a standard gaussion DRAG pulse. (b) Numerical simulation of exact dynamics shown by RWA Hamiltonian of Eq.\ref{3 lvl RWA} for Gaussian and Gaussian DRAG pulse for anharmonicity $\Delta =-450\;\mathrm{MHz}$, peak pulse amplitude $A_{0}=200\;\mathrm{MHz/2\pi}$, pulse width (standard deviation) $\sigma=6.5$ and coupling between $|0\rangle-|1\rangle$ states as $\lambda=\sqrt{2}$. The simulation is carried out using QuTiP in Python.}\label{DRAG}
\end{figure}

This is the Derivative Removal by Adiabatic Gate (DRAG) substitution. DRAG works by introducing a derivative quadrature component that cancels leakage arising from possible off-resonant excitations due to the multilevel structure of the transmon, up to first order.
$A_{-}$ would similarly be zero, and hence any leakage at the first order is removed exactly. 
The second term in $A_{1}$ integrates to $-\frac{\iota}{2}(I(T)-I(0))$, which is also zero due to our pulse shape assumption. Hence,
\begin{equation}
    \Omega_{1}(T) = -\frac{\iota\sigma_{01}^{x}}{2}\Big(\int_{0}^{T}I(t)dt\Big).
    \label{1order final}
\end{equation}

Thus, if the area under the curve of I(t) is $\pi$, we should get the desired X gate (Eq. \ref{1order final}) if the errors arising from higher orders in the expansions are negligible. 
Fig. \ref{DRAG}(a) shows the Gaussian DRAG pulse shape. Fig. \ref{DRAG}(b) shows the improvement of DRAG over a simple plain Gaussian pulse.

It is constructive to see what other errors arise in higher orders. 
Let us consider the commutator of the second order term $\Omega_{2}(t,t_{0})$ ignoring the coefficients that arise from the integration,
\begin{equation}
	\begin{split}
		[H(t_{1}),H(t_{2})]=[H_{01}(t_{1}),H_{01}(t_{2})]+[H_{01}(t_{1}),H_{12}(t_{2})]\\+[H_{12}(t_{1}),H_{01}(t_{2})]+[H_{12}(t_{1}),H_{12}(t_{2})].
	\end{split}\label{magnus order 2}
\end{equation}
The first of the four commutators of the right-hand side term of Eq. \ref{magnus order 2} gives,
\begin{equation}\label{AC01}
		[H_{01}(t_{1}),H_{01}(t_{2})] = P(t_{1},t_{2})\sigma_{01}^{z}.
\end{equation}
The last commutator term in Eq.\ref{magnus order 2} gives,
\begin{equation}\label{AC12}
		[H_{12}(t_{1}),H_{12}(t_{2})] = Q(t_{1},t_{2})\sigma_{12}^{z}.
\end{equation}
The commutator involving $H_{01},H_{12}$ in Eq.\ref{magnus order 2} gives,
\begin{equation}\label{L02}
		[H_{01}(t_{1}),H_{12}(t_{2})]+[H_{12}(t_{1}),H_{01}(t_{2})] = R(t_{1},t_{2})\sigma_{02}^{+}+S(t_{1},t_{2})\sigma_{02}^{-}.
\end{equation}

$P(t_{1},t_{2})$, $Q(t_{1},t_{2})$, $R(t_{1},t_{2})$ and $S(t_{1},t_{2})$ in Eq. \ref{AC01}-\ref{L02} are time-dependent functions involving $\Omega_{I}$, $\Omega_{Q}$, $e^{\pm\iota\Delta t_{1}}$ and $e^{\pm\iota\Delta t_{2}}$. 
The analytical pulse-shaping task is then to identify in-phase and quadrature signals such that the integrals of these functions also become zero in the respective orders in which they appear in the expansion. 
Here, we stop short of giving the explicit form and focus only on the operator and, hence, the error channel that arises. 

We see that the AC-Stark effect-induced phase errors ($\sigma_{k,k+1}^{z}$ in Eq. \ref{AC01}-\ref{AC12}) arise at second order in the expansion. Both the $|0\rangle-|1\rangle$ and $|1\rangle-|2\rangle$ levels are affected; however, the phase error in the $|0\rangle-|1\rangle$ level can be managed by detuning the original drive \cite{motzoi2009simple}.  

Leakage error at the first order was completely eliminated, but there exists another leakage channel (Eq.\ref{L02}) at the second order, which couples the ground state and leakage state directly.
This $|0\rangle-|2\rangle$ coupling at the second order becomes the dominant leakage channel.
The DRAG substitution suppresses this error channel but cannot eliminate it completely.

\begin{tcolorbox}[colback=gray!5!white, colframe=gray!75!black, title= What DRAG is Not]

DRAG is not a one-stop solution to suppress all error channels. It eliminates the dominant first-order leakage error and helps in the suppression of some second-order errors. However, the second-order errors remain, and so do the other higher-order errors. As the design of faster gates becomes important, so does the mitigation of higher-order errors. 
Consequently, a higher-order DRAG analogue protocol becomes necessary. 
\end{tcolorbox}

Designing the pulse by minimizing or eliminating certain orders of expansion is just effective Hamiltonian engineering. While engineering the elimination of even higher-order terms is needed, these often entail high hardware overhead. Further, the analytical exercise of designing pulses for more accurate gate fidelities hinges on the assumption of ideal hardware components. However, errors in hardware components are detrimental to gate operation. Knowledge of the hardware and the error channels arising from them then becomes equally important. We shall see some hardware components in Sec.\ref{sec4}.

\section{Engineering of Pulse Shaping}\label{sec4}
\vspace{0.25cm}
The previous section explored the physics of pulse shaping, focusing on abstract models and idealized dynamics. However, these strategies assume perfect hardware calibration and performance, which is rarely the case. To fully appreciate pulse shaping as a tool for quantum gate engineering, students must also understand the engineering realities of it. By situating pulse shaping within its experimental context, readers gain a comprehensive perspective that motivates both the challenges of implementation and the need for hardware‑aware theoretical developments. 

We try to bridge theory and practice for the reader in this section by:
\begin{enumerate}[label=(\roman*)]
    \item Describing how control pulses are generated and delivered using laboratory hardware.
    \item Identifying common hardware limitations such as distortion, finite sampling, and phase fluctuations.
    \item Analyzing how imperfections in the signal chain map onto physical error channels in qubit dynamics.
\end{enumerate}

Pulse synthesis can be performed using one of the following methods: analog IQ mixing, direct digital synthesis, or digital up-conversion (DUC). 
The first is how control pulses have historically been generated in labs. 
However, the latter methods have emerged as an alternative in recent years, mitigating many of the errors that might otherwise arise from the wide calibration required across an array of analog components.

\begin{figure}[h]
\centering
\includegraphics[scale=0.55]{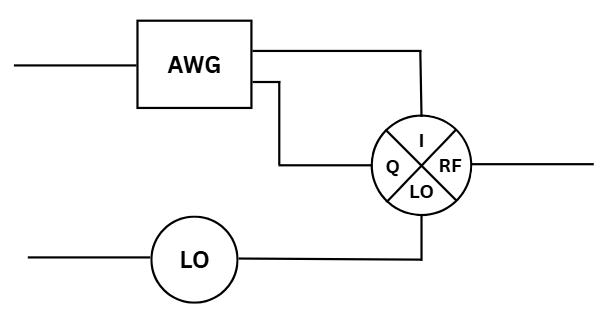}
\caption{\justifying Schematic of the hardware essential for pulse generation. The in-phase (I) and quadrature signals (Q) from the arbitrary waveform generator (AWG) combine with the sinusoid from the local oscillator (LO) in the IQ mixer. }\label{Hardware}
\end{figure}

We first describe the components of the analog IQ-mixing approach, which directly relate to the two-control driving scheme in the Hamiltonian structure. 
Instrument imperfections and miscalibrations cause the breakdown of our DRAG pulse, and these errors are comparable to environmental errors.
A schematic of our signal chain is shown in Fig.\ref{Hardware}

\subsection{Arbitrary Waveform Generator (AWG)}\label{subsec31}

An Arbitrary Waveform Generator (AWG) is an important piece of equipment in the pulse generation signal chain. The AWG is where the "shaping" of the pulse takes place. It is capable of generating any mathematical user-defined pulse shape, a leap beyond function generators, which can generate only basic square and triangular waves. The analytical waveform is defined and stored as a sequence of voltage data points. The digital-to-analog converter (DAC) samples these to generate an analog signal. These signals are then sent to an IQ mixer for analog mixing and up-conversion. 

The pulse, initially defined by the user in the AWG, is defined in the baseband region (near 0 Hz) before it is upconverted. Modern AWGs have sufficiently high sampling rates (over 10 Giga Samples per second) to faithfully reproduce the pulses required for gate operations in quantum computing. Before going any further, let us look at the Nyquist-Shannon sampling theorem. This governs our digital-to-analog conversion process.

\subsubsection{Sampling Theorem and Nyquist Zones}\label{subsubsec311}
\begin{theorem}[Nyquist-Shannon Theorem]\label{thm1}
If a function x(t)contains no frequencies higher than B hertz, then it can be completely determined from its sampled values at a sequence of points spaced less than 1/(2B) seconds apart. 
\end{theorem}

\begin{figure}[h]
\centering
\includegraphics[scale=0.5]{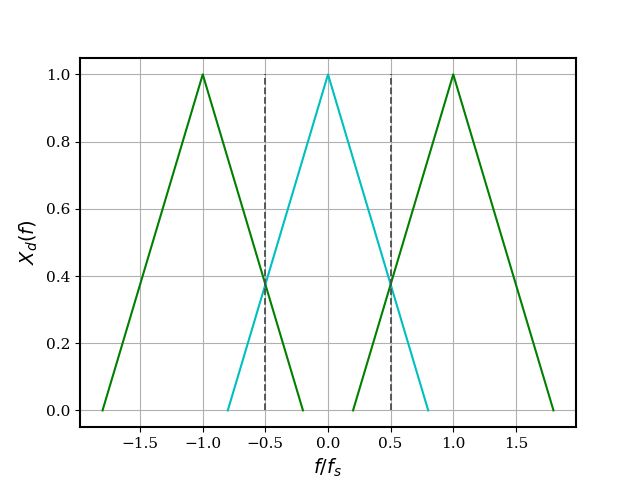}
\caption{\justifying Overlap of the discrete Fourier transform $X_{d}(f)$ images due to improper sampling ($f_{s}\leq2f$).}\label{fig2}
\end{figure}

In other words, for a baseband signal, which is centered around 0 Hz with the highest frequency content $f$, it must be sampled with at least $2f$ sampling rate. This requirement can be understood by taking the discrete Fourier transform $X_{d}(f)$ of the sampled points $x[n]$ sampled every $T_{s}$ seconds,
\begin{equation}
    \begin{split}
        X_{d}(f) = \sum_{n=-\infty}^{\infty}x[n]\,e^{-\iota 2\pi nfT_{s}}, 
    \end{split}
\end{equation}\label{discrete fourier transform 1}
\begin{equation}
    \begin{split}
        X_{d}(f+f_{s}) = X_{d}(f),\qquad T_{s}=\frac{1}{f_{s}}. 
    \end{split}\label{discrete fourier transform 2}
\end{equation}

Clearly, $X_{d}(f)$ is a periodic function, with copies in the frequency domain shifted by the sampling frequency $f_{s}$. For a signal with bandwidth (range of frequency content) greater than $f_{s}/2$, it clearly overlaps with its copies as shown in Fig. \ref{fig2}. Such overlaps cause the frequency components of the signal above $f_{s}/2$ to be indistinguishable from a "partner" lower-frequency component, called an alias, associated with one of the copies. With a sufficiently high sampling rate and a low-pass filter, one can easily obtain the desired signal without aliasing.

\subsubsection{Higher Nyquist-Zones}\label{subsubsec312}
Aliasing is not always bad, and higher Nyquist zones play a role in modern RF generation. One defines the interval $[(m-1)\frac{f_{s}}{2},m\frac{f_{s}}{2}],m\in \mathbb{N}$ as the $m^{th}$ Nyquist zone. In systems that require a signal frequency greater than the sampling rate but limited by it, aliasing is exactly how the signals are generated. Instead of using a low-pass filter, a band-pass filter is used to extract the higher-frequency signal, which lies in the higher Nyquist zone. 

When sampling a sinus waveform of frequency $f_{0}$, it will appear as the fundamental spectral component at $f_{0}$ in the frequency domain. However, there will be additional frequency products, often called images, at higher frequencies. These products depend on the sampling rate. This directly follows from Eq. \ref{discrete fourier transform 2},
\begin{equation}
    \tilde{f} = |C.f_{s}\pm f_{0}|,\quad C\in \mathbb{W}.
\end{equation}\label{image formula}

\begin{figure}[h]
\centering
\includegraphics[width=0.55\textwidth]{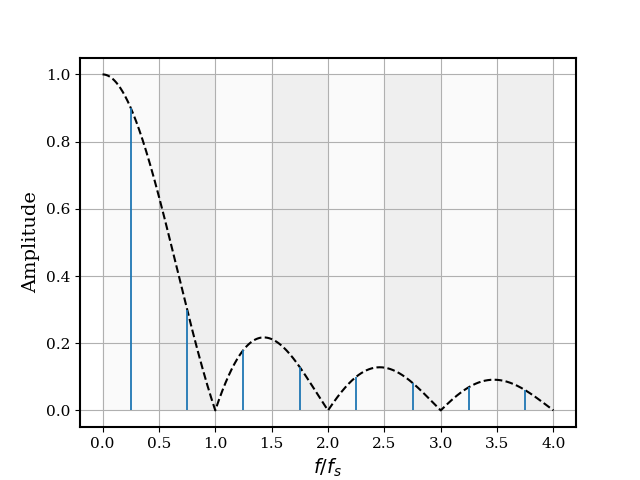}
\caption{\justifying Even though aliasing can be used to generate RF signals, the amplitude shows a $\mathrm{sinc}(\pi x)$ roll-off. The alternating shades of grey denote the different higher Nyquist zones.}\label{Rolloff}
\end{figure}

After sampling the waveform defined in the AWG, the digital signal is sent to the DAC. 
In conventional DACs, the output is a set of discrete voltage levels, each held for $T_{s}$, the sampling period. 
Therefore, the frequency response of such a DAC is of $\mathrm{sinc}(\pi x)=\sin(\pi x)/(\pi x)$ (Fourier transform of the square hold-off characteristic of the voltage levels). 
The amplitude of the images generated at frequencies higher than those in the first Nyquist zone is dependent on the roll-off of $\mathrm{sinc}(\pi x)$. 
So, higher-frequency images are attenuated, as shown in Fig. \ref {Rolloff}.

\subsubsection{Consequence of finite sampling}\label{subsubsec313}

This finite sampling rate has a direct consequence for the pulse-shaping
strategy developed in Sec.~2.3. 
The DRAG condition of Eq.~\eqref{DRAG Substitution} requires the quadrature component $Q(t)$ to track the exact time derivative of the in-phase component $I(t)$.
Since differentiation amplifies high-frequency content, $Q(t)$ generally carries more spectral weight near the Nyquist frequency $f_s/2$ than $I(t)$ does for a pulse of the same duration. 
If the AWG's sampling rate is insufficient relative to the pulse bandwidth, or if the DAC's $\mathrm{sinc}(\pi x)$ roll-off (Fig.~\ref{Rolloff}) attenuates this high-frequency content unevenly between the $I$ and $Q$ channels, the digitally synthesized quadrature no longer satisfies Eq.~\eqref{DRAG Substitution} exactly. \
The cancellation in Eq.~\eqref{integration} is then incomplete, and the residual rotation error $\sigma^y_{01}$ and leakage terms $\sigma^{\pm}_{12}$ that DRAG is designed to remove reappear at a level set by the AWG's sampling rate relative to the pulse bandwidth, independent of how carefully the analytical pulse shape itself was designed. 
This is one concrete reason modern AWGs are specified with sampling rates far in excess of the bandwidth implied by the gate's pulse duration alone, as discussed in Sec.~\ref{subsec31}

\subsection {Local Oscillator (LO)}\label{subsec43}

The local oscillator is a device used in the mixing process which, changes the frequency of the input signal. In conjunction with a mixer it outputs the sum $f_{LO} + f_{I}$ and the difference $f_{LO} - f_{I}$ of the local oscillator frequency $f_{LO}$ and the input frequency $f_{I}$. Crystal oscillators provide a low-cost, stable LO; however, they are single-frequency devices. A variable-frequency oscillator is used to generate a range of frequencies, but at the expense of phase and frequency stability. This is where a phase-locked loop (PLL) comes in.

\subsubsection{Phase locked loop (PLL)}\label{subsubsec321}
\begin{figure}[h]
\centering
\includegraphics[scale=0.5]{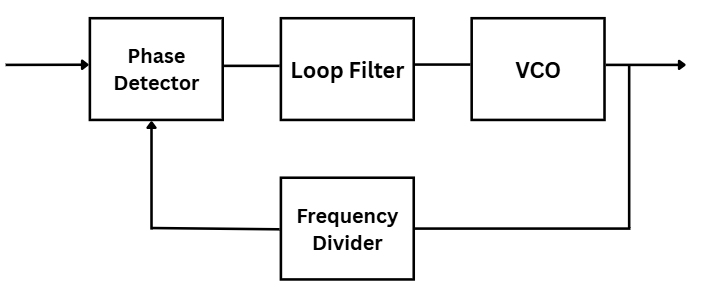}
\caption{Schematic of a basic phase-locked loop configuration.}\label{PLL}
\end{figure}

A phase-locked loop is a system that generates an output signal, which is fed back into the loop and constantly compared to a reference signal. The input and the output signal then have a fixed constant relative phase relation. PLL typically has a Phase detector, a Loop filter, a Voltage-controlled oscillator (VCO), and a Frequency divider, as shown in Fig. \ref{PLL}. The key part tied to the local oscillator is the VCO, which is often the local oscillator itself. The local oscillator output (which may be a VCO) feeds back into the PLL. The PLL compares its phase/frequency to a stable reference (usually a crystal oscillator). The PLL adjusts the control voltage to the VCO/LO. The local oscillator output becomes stable, low-drift, and tunable.

\subsubsection{Consequence of phase fluctuations}\label{subsubsec322}
The stability of phase and frequency is particularly important. Random fluctuations lead to qubit dephasing. These errors are comparable to environmental dephasing. So, the very instrument used to manipulate qubit information can render the qubit useless if not calibrated or handled carefully. Note that the qubit's phase is set by reference to the rotating frame, which in turn is set by the LO. If LO is unstable, so is the rotating frame, and the phase coherence of the qubit is lost due to random fluctuations \cite{ball2016role}. Consider the following Hamiltonian in $\hbar=1$ units,
\begin{equation}
    H = \frac{1}{2}\omega_{0}\sigma_{z} + \frac{1}{2}\delta\omega_{0}(t)\sigma_{z} + A(t)\cos(\omega_{\mathrm{LO}}t+[\phi_{C}(t)+\phi_{N}(t)])\sigma_{x}.
\end{equation}\label{LO Hamiltonian}

The first term describes the free evolution with the qubit's resonant frequency $\omega_{0}$. The second term accounts for contributions from environmental noise, which shifts the transition frequency to $\omega_{0} \xrightarrow{}\omega_{0}+\delta\omega_{0}(t)$. The third term is the time-dependent interaction term with LO driving the system at carrier frequency $\omega_{\mathrm{LO}}$ and the drive amplitude being $A(t)$. The phase of the drive has two parts: the control phase $\phi_{C}(t)$, which dictates the desired phase evolution, and the noise part $\phi_{N}(t)$, which is the random fluctuations. Now, we apply two interaction-picture transformations. First, we rotate with the carrier frequency and make the rotating wave approximation to get,
\begin{equation}
    \begin{split}
        H_{1} &= \frac{1}{2}(\omega_{0}-\omega_{\mathrm{LO}})\sigma_{z} + \frac{1}{2}\delta\omega_{0}(t)\sigma_{z}+ \frac{1}{4}A(t)\{e^{-\iota[\phi_{C}(t)+\phi_{N}(t)]}\sigma^{+}+e^{+\iota[\phi_{C}(t)+\phi_{N}(t)]}\sigma^{-}\}.
    \end{split}
\end{equation}\label{LO interaction 1}

The $\sigma^{\pm}$ operators are the ladder operators.
Now, the instantaneous rate of change of the phase fluctuations is nothing but the fluctuations of the carrier frequency, so $\delta\omega_{\mathrm{LO}}(t)=\dot{\phi}_{N}(t)$. This means that phase fluctuations in LO lead to $\omega_{\mathrm{LO}} \xrightarrow{}\omega_{\mathrm{LO}}+\delta\omega_{\mathrm{LO}}(t)$.
Now, implementing another transformation such that the unitary transform is defined as $V=\exp(\iota\frac{\phi_{N}(t)}{2}\sigma_{z})$. The goal is to move into this noisy frame to see what the drive's phase noise really does to the system.
As it happens, the unitary is nothing but the evolution operator under the dephasing Hamiltonian arising due to fluctuations in the LO, $H_{\phi}=\frac{1}{2}\dot{\phi}_{N}(t)\sigma_{z}$. Setting the static detuning ($\omega_{0}-\omega_{\mathrm{LO}}$) to zero, the overall transformed Hamiltonian then becomes,
\begin{equation}\label{LO interaction 2}
    H_{2} = \frac{1}{2}\delta\omega_{0}(t)\sigma_{z} +\frac{1}{2}\dot{\phi}_{N}(t)\sigma_{z} + \frac{1}{2}A(t)\{\cos[\phi_{C}(t)]\sigma_{x}+\sin[\phi_{C}(t)]\sigma_{y}\}.
\end{equation}

One can clearly see that the second term, dephasing from LO, is comparable to the first term, the environmental dephasing term. Hence, LO drift directly leads to qubit dephasing. So, hardware errors have a clear physical effect on the system.

\subsection {IQ Mixing}\label{subsec44}

The in-phase and quadrature (IQ) mixer setup combines components from the AWG and the LO to generate the pulse. The superconducting qubit's resonant frequency lies in the GHz range. So, it is necessary to upconvert our baseband pulse signal, which lies near 0Hz, to the GHz range. The mixing process generally produces two signals after upconversion: one at LO+IF and one at LO-IF, where IF (intermediate frequency) is the baseband signal's frequency. The undesired signal is often called the image. While the upconversion process can be implemented using various mixing techniques, IQ mixing reduces the need for multiple filters and helps reject the image signal. Further, the two controls, in-phase (I) and quadrature (Q), help define a more general, arbitrary pulse or signal.

\subsubsection{Mapping control knobs to the Hamiltonian}\label{subsubsec331}
The I and Q signal channels act like the control knobs for the $\sigma_{x}$ and $\sigma_{y}$ terms defined relative to our system Hamiltonian. 
Changing phase across these channels implies changing the qubit rotation axis. In this analogy, a Z gate is just a phase update of the I and Q channels. This can be understood as, if we change the phase $\phi\xrightarrow{}\phi+\pi/2$, then $I\xrightarrow{}Q$ and $Q\xrightarrow{}-I$, and consequently the rotation axis changes. At the same time, the application of $\sigma_{z}$ on $\sigma_{x}$ and $\sigma_{y}$ yields $\sigma_{z}\sigma_{x}=\iota\sigma_{y}$ and $\sigma_{z}\sigma_{y}=-\iota\sigma_{x}$. So, the application of the Z gate and the change in phase across the I and Q channels are equivalent. So, we do not need a pulse for the Z gate, and it is, as aptly noted, known as the virtual Z gate.

Assuming everything is perfect, we now have the perfect pulse $S(t)$ ready for qubit manipulation,
\begin{equation}\label{IQ mixing 1}
    S(t)=\Re\{(I+\iota Q)\,\exp(\iota \omega_{\mathrm{LO}}t)\},
\end{equation}
the area of the complex baseband holds the information about qubit rotation, so changing the area changes the amount of rotation applied. 
Under the DRAG pulse condition, the rotation is controlled just by the in-phase (I) baseband.

\subsubsection{Consequence of imperfect mixing}\label{subsubsec332}
Any voltage fluctuations, gain mismatches, or DAC scaling mismatches in the I and Q channels result in amplitude errors, 
\begin{equation}
    \begin{split}
        \tilde{S}(t) &= I(t)\cos(\omega_{\mathrm{LO}}t)-Q(t)\sin(\omega_{\mathrm{LO}}t+\phi)\\
        &=I(t)\cos(\omega_{\mathrm{LO}}t)-Q(t)\sin(\omega_{LO}t)\cos(\phi)-Q(t)\cos(\omega_{\mathrm{LO}}t)\sin(\phi)\\
        &=\{I(t)-Q(t)\sin(\phi)\}\cos(\omega_{\mathrm{LO}}t)-\{Q(t)\cos(\phi)\}\sin(\omega_{\mathrm{LO}}t)\\
        &=\tilde{I}(t)\cos(\omega_{\mathrm{LO}}t)-\tilde{Q}(t)\sin(\omega_{\mathrm{LO}}t).
    \end{split}
\end{equation}\label{IQ Skew}

The DRAG pulse design hinges on the perfect orthogonality of the I and Q signals. However, a skew that breaks perfect orthogonality leads to imperfect mixing of the baseband signal. The imperfections give rise to fluctuations and distortion in the pulse shape, resulting in over- and under-rotations of the qubit. This distortion is also detrimental to mitigating qubit leakage errors. The skew might be due to asymmetry in the LO used for the I and Q paths, or as simple as different cable lengths leading to varying path lengths for the I and Q channels. There exist other ways, like mixer compression, where the pulse shape can get distorted.

\begin{figure}[h]
\centering
\includegraphics[width=0.7\textwidth]{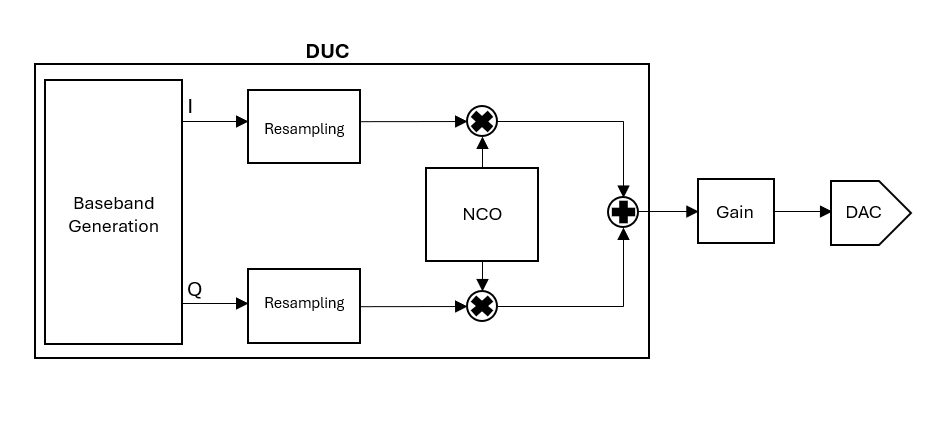}
\caption{\justifying Schematic of the digital up-conversion process. The complete mixing process is done digitally. The final digital waveform is then sent to the DAC for analog pulse signal generation.}\label{DUC}
\end{figure}
\subsection {Digital Up-Conversion (DUC)}\label{subsec45}
While the analog setup has vast applicability across diverse fields, the digital up-conversion (DUC) process has become the modern standard. This shift from the analog setup is due to the elimination of errors generated there, such as LO leakage and IQ skew. In the DUC process, the in-phase and quadrature baseband signals are resampled and then scaled by the numerically controlled oscillator (NCO). 

The use of NCO provides finer frequency resolution, supports variable frequencies, and enables switching between them at very high speeds. Further, any phase or frequency drift relative to the I and Q channels is also minimised. The complete digitization of the process results in extremely accurate matching of the I and Q signals, preventing quadrature imbalance and maintaining orthogonality between the two. The existence of a sinc-rolloff of amplitude in Fig.\ref{Rolloff} is compensated by an inverse sinc gain after the DUC process. Only after this complete digitization process does the signal go to a DAC, as shown in Fig.\ref{DUC}.

\section{Two-Qubit Gate}\label{sec5}
\vspace{0.25cm}
Having examined single qubit pulse shaping and its engineering realities, we now turn to the more complex task of two qubit gate implementation. Two-qubit gates are the gateway to entanglement generation, which is the cornerstone of universal quantum computation. Implementation of two-qubit gates then becomes central to various algorithmic routines, yet they are far more sensitive to both theoretical design and hardware imperfections. It is then imperative to study how the underlying pulse implementation differs from the single-qubit case. 

Often, the CNOT gate is the first two-qubit gate introduced in introductory quantum information lectures \cite{nielsen2010quantum}. However, the underlying hardware architecture of quantum computing yields distinct native gate sets. One then, with the help of both single and two-qubit native gate sets, mimics the operation of a CNOT gate \cite{superconducting_review}. In particular, the cross resonance (CR) gate provides a natural platform for fixed capacitively coupled superconducting qubits, but its dynamics intertwine control pulses, qubit detuning, and unwanted interactions.

Unlike the approach we took in Sec. \ref{sec3} to explicitly derive the Hamiltonian, the two-qubit gate case becomes messy, and presenting it here would interupt the pedagogical narrative. We emphazize on the conceptual clarity rather than mathematical rigour in this section. Readers can look into Ref.\cite{magesan2020effective,Effective_Hamiltonian_Derivation} for explicit details on the effective Hamiltonian derivation.

In this section we take the reader through,
\begin{enumerate}[label=(\roman*)]
    \item The decomposition of the CNOT gate into the CR gate and single-qubit gates.
    \item We directly start by describing the effective Hamiltonian and provide an intuitive explanation for the form of the Hamiltonian.
    \item Relate the Hamiltonian terms to the various physical error channels.
    \item Review the pulse-shaping and sequencing strategies used to mitigate these errors.

\end{enumerate}

\subsection{Cross-Resonance (CR) Gate}\label{subsec51}
The $\mathrm{CR}$ gate is locally equivalent to the CNOT gate. That is, we need to apply only a few single-qubit gates to make its operation equivalent to that of the CNOT gate. As in Eq.\ref{ZX Matrix}, we define $Z\otimes I_{\theta}\equiv \exp{(-\iota\frac{\theta}{2} Z\otimes I)}$ and $I\otimes X_{\theta}\equiv \exp{(-\iota\frac{\theta}{2} I\otimes X)}$, 

\renewcommand{\arraystretch}{0.65}
\begin{equation}
    \begin{split}
        \mathrm{CR}_{\theta}\equiv Z\otimes X_{\theta}\equiv \exp{(-\iota\frac{\theta}{2} Z\otimes X)}= \begin{pmatrix}
            \cos{\frac{\theta}{2}}& \;-\iota\sin{\frac{\theta}{2}}&0&0\\
            \\
            -\iota\sin{\frac{\theta}{2}}& \;\cos{\frac{\theta}{2}}&0&0\\
            \\
            0&\;0&\;\cos{\frac{\theta}{2}}& \;\iota\sin{\frac{\theta}{2}}\\
            \\
            0&0&\iota\sin{\frac{\theta}{2}}& \;\cos{\frac{\theta}{2}}
            \end{pmatrix}.
    \end{split}\label{ZX Matrix}
\end{equation}

\begin{tcolorbox}[colback=gray!5!white, colframe=gray!75!black, title= An Intuitive Analogy]
Consider the following intuitive situation. An animal handler has two parrots who have been trained to obey certain commands. The two parrots, A and B, only respond to their own respective commands. Now, the handler whispers in Parrot A’s ear, “Parrot B, turn.” The handler’s command does not affect parrot A. However, parrot A vocalised the command, which parrot B hears and turns. Further, depending upon parrot A’s internal state, being calm or agitated, it vocalises immediately or with some delay, and parrot B responds appropriately.

The different commands for different parrots are like different qubit transition frequencies for different sets of qubits. The parrots hearing each other is the coupling between the qubits. The handler issuing the command is analogous to driving the qubit. So, just as the parrot example creates a conditional response from parrot B (immediate or delayed turn) depending on parrot A (calm or idle). Driving one of the qubits (control) at the transition frequency of the other qubit (target) results in the latter to oscillate coherently between the $|0\rangle$ and $|1\rangle$ levels. The rate of such evolution depends on the state of the control qubit; a $Z\otimes X$ interaction is induced, thus enabling an entangling operation in the two-qubit subspace.
\end{tcolorbox}

Upto to a global phase factor of $e^{-\iota\frac{\pi}{4}}$, the CNOT is implemented by the following gate sequence \cite{superconducting_review},
\begin{equation}\label{CNOT decomposition}
        \mathrm{CNOT} \equiv e^{-\iota\frac{\pi}{4}}(Z\otimes I_{\frac{\pi}{2}})(I\otimes X_{\frac{\pi}{2}})(\mathrm{CR}_{-\frac{\pi}{2}}).    
\end{equation}

\begin{figure}[h]\label{CNOTCR}
\centering
\includegraphics[scale=0.5]{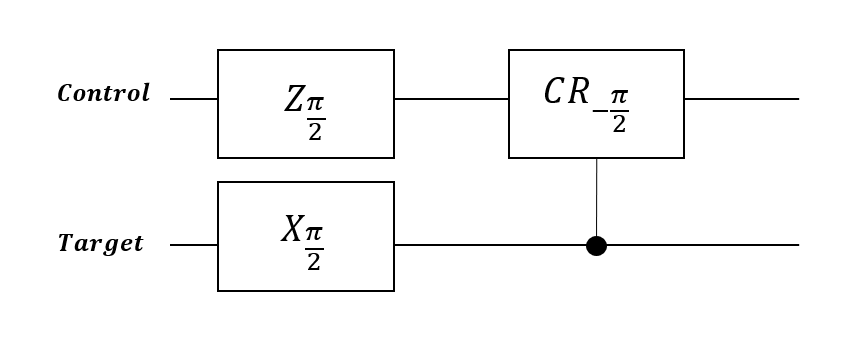}
\caption{\justifying Decomposition of CNOT gate into the cross-resonance gate and single-qubit gates.}
\end{figure}

\begin{tcolorbox}[colback=gray!5!white, colframe=gray!75!black, title= Asymmetry of control and target qubit]
On the surface, it may seem like interchanging the drive on two different qubit interchanges the control and target qubit. However, by the principal mechanism of the CR gate, the two qubits' transition frequency or the energy level spacing between $|0\rangle-|1\rangle$ is required to be different. So, to flip the control and target requires updating the drive frequencies as well and not just switching the drive onto the other qubit.
\end{tcolorbox}

\subsection{Effective CR Hamiltonian and Echo Sequence}\label{subsec52}
The effective CR Hamiltonian reveals an important distinction from the single-qubit case, 
\begin{equation}
    \begin{split}
        H_{\mathrm{CR}_{\mathrm{eff}}}(\Omega) &= \omega_{ix}\frac{I\otimes X}{2} + \omega_{iz}\frac{I\otimes Z}{2} + \omega_{zi}\frac{Z\otimes I}{2}+ \omega_{zx}\frac{Z\otimes X}{2} + \omega_{zz}\frac{Z\otimes Z}{2},
    \end{split}\label{CR Hamiltonian}
\end{equation}
Even without considering possible leakage to higher excited states, the Hamiltonian contains additional terms beyond the desired $Z\otimes X$ term \cite{Effective_Hamiltonian_Derivation}. The intuition of why this must be the case can be understood by again considering the animal handler and the two-parrot scenario. When the handler communicates to parrot A, parrot B is not in isolation, so it may overhear the handler, leading it to turn anyway ($I\otimes X$ term). Both the parrots have their own internal states ($I\otimes Z$ and $Z\otimes I$ terms). Further, one of the parrots may influence the other parrot's mood. After all, parrots themselves communicate ($ Z\otimes Z$ term).

The presence of these extra terms leads to errors in the computational subspace, thereby reducing fidelity. These other terms arise due to various reasons, from the control qubit experiencing AC Stark shift from the off-resonant drive to the CR drive leaking into the target qubit and enabling resonant Rabi oscillation and AC Stark shift on the target qubit. Further, these terms are not on equal footing. They differ by orders of magnitude. This is encoded in their coefficients $\omega_{ab}$ via their dependence on the drive amplitude $\Omega$ and qubit-qubit coupling $J$, as in Eq. \ref{Coefficient relation}. The CR gate is implemented in the $J<<\Omega$ regime,
\begin{equation}
    \begin{split}
        \omega_{ix} \propto J\Omega,\quad
        \omega_{zx} \propto J\Omega,\quad
        \omega_{zz} \propto J^{2},
        \omega_{zi} \propto \Omega^{2},\qquad
        \omega_{iz} \propto J^{2}.\qquad
    \end{split}\label{Coefficient relation}
\end{equation}

A single pulse cannot counteract the additional terms and implement the ideal CR gate. We need an alternative strategy. Instead of focusing on a singular pulse, we try combining a sequence of pulses. The goal would be to determine the optimal pulse shapes and sequence, while keeping the number of pulses to a minimum. The very first strategy is to implement an echo sequence \cite{Effective_Hamiltonian_Derivation}. It consists of a four-pulse sequence. First, a CR pulse with half the ideal duration required for implementation is applied, followed by a $\pi$-pulse on the control. A CR pulse is then repeated, albeit with negative amplitude (the Hamiltonian is given by Eq.\ref{CR Hamiltonian II}),
\begin{equation}
    \begin{split}
        H_{CR_{eff}}(-\Omega) &= -\omega_{ix}\frac{I\otimes X}{2} + \omega_{iz}\frac{I\otimes Z}{2} + \omega_{zi}\frac{Z\otimes I}{2}- \omega_{zx}\frac{Z\otimes X}{2} + \omega_{zz}\frac{Z\otimes Z}{2},
    \end{split}\label{CR Hamiltonian II}
\end{equation}
followed by another $\pi$-pulse (Eq.\ref{Echo Unitary I}). This scheme removes the $Z\otimes Z$, $Z\otimes I$, and $I\otimes X$ term effects from the target qubit dynamics. The resulting unitary is a linear combination of the desired $Z\otimes X$ term and other minute residual error terms as shown in Eq. \ref{Echo Unitary II},
\begin{equation}
    \begin{split}
        U_{\mathrm{echo}}(\Omega,\frac{\tau}{2})=R_{X}(-\pi)U_{\mathrm{CR}}(-\Omega,\frac{\tau}{2})R_{X}(\pi)U_{\mathrm{CR}}(\Omega,\frac{\tau}{2}),
     \end{split}\label{Echo Unitary I}
\end{equation}
\begin{equation}
    \begin{split}
        U_{\mathrm{echo}}(\Omega,\frac{\tau}{2})= a_{ii}I\otimes I + a_{iy}I\otimes Y + a_{iz}I\otimes Z+a_{zx}Z\otimes X.
    \end{split}\label{Echo Unitary II}
\end{equation}
\begin{figure}[h]
\centering
\includegraphics[scale=0.6]{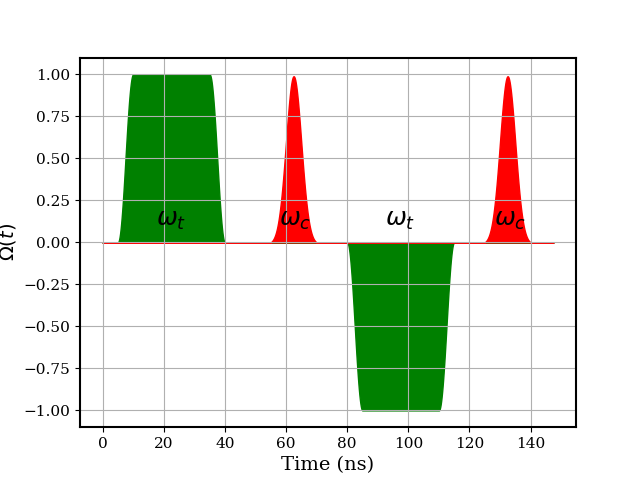}
\caption{Schematic of a cross-resonance gate echo sequence. The CR drive (green) pulse is a flat top Gaussian with frequency of the target qubit $\omega_{t}$. Echo sequence also consists of $\pi$-pulses (red) on the control with its own characteristic transition frequency $\omega_{c}$. }\label{Echo CR}
\end{figure}
Using a flat-top Gaussian for the CR drive (as the effective Hamiltonian derivation assumes a near constant drive amplitude) and a Gaussian-DRAG pulse for the single qubit rotation for the sequence (Fig.\ref{Echo CR}), one can implement a higher fidelity CR gate compared to a single pulse implementation.

\subsection{Active Cancellation}\label{subsec53}
\begin{figure}[h]
\centering
\includegraphics[scale=0.65]{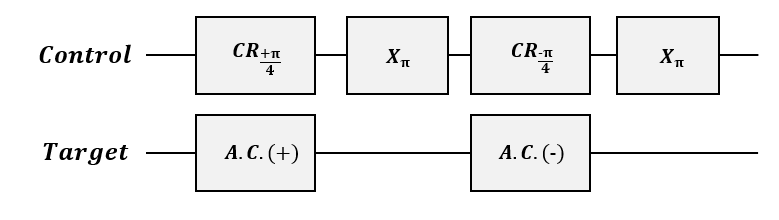}
\caption{\justifying Schematic of an echoed CR gate sequence with the addition of a simultaneous active cancellation (A.C.$\pm$) pulse. The first A.C. pulse is positive in amplitude, and the second is negative.}\label{Active}
\end{figure}

While the echo sequence provided a significant improvement, the fidelity can still be improved by leaps and bounds. The active cancellation method prescribes adding an additional pulse to the target qubit simultaneously with the CR drive pulse, as shown in Fig. \ref {Active}. The active cancellation protocol counteracts the $I\otimes X$ and $I\otimes Y$ terms. The exact parameters required to drive the new pulse scheme are determined by a systematic calibration measurement procedure as introduced in Ref.\cite{Active_Cancellation}. A series of Rabi oscillation experiments is performed to measure the Hamiltonian parameters as a function of the CR drive amplitude and phase. It is important to note that the presence of a $I\otimes Y$ term in the effective Hamiltonian renders the echo procedure not as effective, failing to refocus the undesired terms.

The prescription to achieve a high-fidelity CR gate is then as follows: first, the CR drive is driven with a phase $\phi_{0}$ such that the $Z\otimes Y$ interaction strength is negligible. This information is gained from the calibration procedure. A phase $\phi_{1}$ is figured out where the $I\otimes Y$ interaction strength drops to zero. Second, if the target qubit drive is driven with a phase $\phi_{1}-\phi_{0}$ with a suitable amplitude, which can be found out by doing amplitude sweeps over a wide range, then the effects of $I\otimes X$ and $I\otimes Y$ terms also drop off. This scheme gives us a near-ideal $Z \otimes X$ interaction.

Such a scheme improves fidelity to 0.99 and enables faster entanglement generation. In addition, a reduction in gate time from 400 ns to around 160 ns becomes possible.
The $Z\otimes Z$ errors in the effective Hamiltonian of participating qubits are taken care of, but the nature of always-on coupling implies $Z\otimes Z$ type errors originating from neighbouring spectator qubits as well. This error channel is addressed by introducing a continuous rotary tone rather than an active-cancellation pulse \cite{Rotary_Tone}. 
The procedure improves upon the idea of active cancellation by taking care of the $I\otimes Y$ term in the echoed unitary and spectator qubit $Z\otimes Z$ error contribution simultaneously.

\subsection{Multi-Derivative DRAG CR Pulse}\label{subsec54}

While the rotary tone method in the aforementioned section does help increase fidelity, the gate times remain around 160 ns. At the same time, calibration requires much larger-amplitude sweeps and multiple-parameter fitting. Reducing gate times increases the likelihood of errors due to non-adiabatic dynamics and increases leakage. For a successful implementation of the CR gate, it is necessary that at the end of the CR drive, the control qubit state remains in the same state, so even the $|0\rangle \leftrightarrow |1\rangle$ transition needs to be suppressed in addition to out-of-computational subspace leakage. 

To circumvent these problems, the multi-derivative CR pulse has been proposed \cite{li2024experimental}. It is a recursive DRAG procedure applied to the CR drive, and a singular DRAG correction on the active cancellation target pulse. 
This procedure alone helps in reducing gate times by enabling the reduction of pulse ramping up time (and ramping down time) from 28 ns to 10 ns, saving around 35 ns in total. 
The increased control precision allows the possibility of increasing the driving amplitude to shorten the holding period, thereby possibly reducing gate time even further. 
All of this is accompanied by the fidelity achieved increasing to around 0.997 \cite{li2024experimental}.  

The rationale behind this recursive approach is that a singular DRAG correction on the CR pulse only corrects a single transition. One can layer corrections to progressively cancel errors. The recursive approach successfully suppresses multiple transitions simultaneously. The recursive expression for the CR drive pulse is as follows: 

\begin{equation}\label{recurssive expression 1}
        \Omega_{\mathrm{CR}} = \Omega_{1} -\iota\frac{\dot{\Omega}_{1}}{\Delta_{10}},
\end{equation}
\begin{equation}\label{recurssive expression 2}
        \Omega_{1} = \Omega_{2} -\iota\frac{\dot{\Omega}_{2}}{\Delta_{21}},
\end{equation}
\begin{equation}\label{recurssive expression 3}
        \Omega_{2} = \sqrt{\Omega_{3}^{2} -\iota\frac{2\Omega_{3}\dot{\Omega}_{3}}{\Delta_{20}}}.
\end{equation}
 The functional forms of Eq. \ref{recurssive expression 1} and Eq.\ref{recurssive expression 2} are just the DRAG condition applied onto the neighbouring coupled levels. The substitution shown in Eq.\ref{recurssive expression 3} relates to the two-photon transition between $|0\rangle$ and $|2\rangle$ levels. Eq.\ref{recurssive expression 3} is valid under the approximation that the pulse-ramp is quasi-adiabatic. Here, $\Omega_{3}$ is the flat top Gaussian pulse. More generally, $\Omega_{3}$ needs to be a continuous pulse, with it being zero at its start and end times. 

Further, it is assumed that active cancellation calibration has been performed and that the Hamiltonian parameters are known. Then, instead of the rotary tone, one adds a DRAG correction to the active cancellation pulse. The rationale behind this is that the phase of the CR drive and the active cancellation together remove the $Z\otimes Y$, $I\otimes X$, and $I\otimes Y$ from the effective CR Hamiltonian,  
\begin{equation}
    \begin{split}
        H^{'}_{\mathrm{CR}_{\mathrm{eff}}}(\Omega) &= u_{iz}\frac{I\otimes Z}{2} + u_{zx}\frac{Z\otimes X}{2} + u_{zz}\frac{Z\otimes Z}{2}.
    \end{split}
\end{equation}\label{new effective hamiltonian}
The only error terms that do not commute with $Z\otimes X$ are $Z\otimes Z$ and $I\otimes Z$. The $Z\otimes X$ and $Z\otimes Z$ terms are related by a $I\otimes Y$ term. A compensatory drive in the quadrature proportional to the derivative, i.e., DRAG pulse, appears to remove the $Z\otimes Z$ term. The $Z\otimes Z$ contribution is rather small, so the amplitude of the DRAG pulse is also of similar strength. The calibration needed to find this is much simpler than the rotary tone, requiring only a few sampling points and linear fitting to find the optimal amplitude. The remaining $I\otimes Z$ term is addressed by slightly detuning the CR drive.
\subsection{Identity Gate to Mitigate Idle-Time Crosstalk}\label{subsec55}
All of the discussion up to this point has been on the qubits that are active in the dynamics. However, as noted earlier, the CR gate is supported by an always-on coupling. This means that two qubits, which are spectators, themselves experience a static $Z\otimes Z$ interaction. Consider the action of $Z\otimes Z$ on the following two speactator states, $|\Psi_{1}\rangle=\frac{1}{\sqrt{2}}(|00\rangle+|11\rangle)$ and $|\Psi_{2}\rangle=\frac{1}{\sqrt{2}}(|00\rangle+|01\rangle)$,

\begin{equation}\label{ZZ 1}
    (Z\otimes Z) |\Psi_{1}\rangle=\frac{1}{\sqrt{2}}(|00\rangle+|11\rangle) = |\Psi_{1}\rangle,
\end{equation}
\begin{equation}\label{ZZ 2}
    (Z\otimes Z) |\Psi_{2}\rangle=\frac{1}{\sqrt{2}}(|00\rangle-|01\rangle)\neq |\Psi_{2}\rangle.
\end{equation}

The always-on coupling leads to the accumulation of different phases across different states during their idle time. A way to mitigate this is to implement a $X^{2}(\pi)\equiv I$ gate on one of the qubits \cite{vandersypen2004nmr}. If the idle time for spectator qubits is $2\tau$, then the spectator qubits evolve under $U(2\tau)=exp(-\iota 2\tau J Z\otimes Z)$. However,
\begin{equation}
    \begin{split}
        X(\pi)U(\tau)X(\pi)=U(-\tau)=U^{\dagger}(\tau).
    \end{split}\label{I gate}
\end{equation}
Eq.\ref{I gate} suggests that applying two $\pi$ pulses on any one of the qubits after time $\tau$ removes the effect of time evolution experienced for time $\tau$ from the always-on coupling. Implementing the identity pulse on a pair of qubits serves as a reset from the disparate phase accumulation that occurs right before a gate operation related to the target algorithm is applied.  


\section{Discussion}\label{sec6}
\vspace{0.25cm}
\subsection{Summary}
In this work, we have presented a pedagogical overview of microwave pulse shaping for superconducting qubits, aiming to bring together physicists and engineers for a combined discussion of abstract Hamiltonian descriptions and their practical implementation on experimental platforms. Starting from the simple observations about simple pulse envelopes, we motivated the need for more refined control techniques by highlighting the role of spectral leakage in weakly anharmonic systems such as the transmon.

The discussion of pulse shaping illustrated how time-domain smoothness translates into reduced spectral broadening, but also emphasized that such approaches alone are insufficient to suppress leakage fully. The need for a better pulse shaping strategy naturally led to the introduction of the derivative removal by adiabatic gate (DRAG) technique, in which the inclusion of a quadrature component proportional to the time derivative of the pulse envelope systematically cancels unwanted off-resonant transitions. Interpreting DRAG through the lens of the Magnus expansion further motivates a more general Hamiltonian engineering perspective, showing how higher-order time-dependent control corrections can be engineered to suppress errors.

\begin{table}[h!]
\centering
\renewcommand{\arraystretch}{1.3}
\begin{tabular}{ |c|c|c|c| } 
\hline
Error Type & Effect& Cause & Hamiltonian Term \\
\hline
\multirow{3}{12em}{\centering Drive induced} & Leakage& Coupling to leakage state & $\sigma_{12}^{\pm}\;(\mathrm{1st \;order})$,$\sigma_{02}^{\pm}\;(\mathrm{2nd \;order})$  \\ 
& Undesired rotations& Residual transverse term & $\sigma_{01}^{y}\;\mathrm{(1st\;order)}$ \\ 
& Phase error & AC Stark effect& $\sigma_{01}^{z},\sigma_{12}^{z}\;(\mathrm{2nd \;order})$ \\
\hline
\multirow{3}{12em}{\centering Hardware and underlying architecture} & Dephasing& Unstable LO & $\sigma_{z}$ \\ 
& Qubit cross-talk& Always on coupling & $\sigma_{z}\otimes\sigma_{z}$ \\
& Pulse distortion & IQ non-orthogonality & - \\
\hline
\end{tabular}
\caption{\justifying The various errors encountered in the discussion can be broadly labeled into two categories of drive-induced errors, and hardware and architecture errors. The cause and effect and the operators with which they appear in the Hamiltonian are summarized here. There also exist environmental errors, but they are beyond the scope of this discussion.}\label{Summary Table}
\end{table}

In practice, due to hardware limitations, control pulses are generated with imperfections, including finite sampling rates, phase discontinuities, and IQ mixer imbalances. Hardware errors introduce additional deviations from the idealized control fields. By examining these effects, we have emphasized that accurate qubit control requires not only careful pulse design but also a detailed understanding of the signal generation chain. A summary of some of the errors encountered in this article's discussion is given in Table \ref{Summary Table}.

Finally, we extended our discussion to multi-qubit operations, particularly the cross-resonance gate. The emergence of effective interactions, along with accompanying unwanted terms, reflects the same interplay between desired system dynamics and multiple error channels encountered at the single-qubit level. Such observations reinforce the broader message that gate design, calibration, and hardware constraints must be understood as a unified problem.

\subsection{Pedagogical implications}
The central pedagogical contribution is not a new pulse-shaping protocol, but a unified narrative that reconstructs pulse shaping from first principles, following the sequence of questions that naturally arise as students transition from textbook quantum gates to experimentally implemented quantum control. By connecting physical intuition, analytical derivations, and practical hardware considerations within a single framework, the article aims to make pulse shaping understandable as a logical consequence of quantum dynamics rather than as a collection of independent control recipes.

This article is written as a bridging text rather than a conventional research review. Its intended reader has completed a first course in quantum mechanics together with an introductory course in quantum computing or quantum information in which qubits and gates are treated as abstract unitaries, largely independent of any physical realization. The natural setting is a follow-on special topics course or an elective bridging quantum information theory and experimental quantum hardware, such as quantum engineering, a guided reading group, or self-directed study by a graduate student entering an experimental superconducting-qubit research group. \\
\vspace{0.25cm}

\subsection{Outlook}
This paper aims to provide an accessible entry point to the practical aspects of quantum control in superconducting circuits. By combining intuitive arguments, analytical tools such as the Magnus expansion, and a discussion of hardware-level effects, we hope to offer a framework that helps readers connect theoretical models with experimental implementations. 

Further, the unified framework of Hamiltonian engineering and hardware effects can be extended to any choice of quantum computing platforms, inspiring unique pulse-shaping strategies relevant for particular choices of platforms. Such a perspective is essential for the development and optimization of high-fidelity quantum operations in current and future quantum processors.

\ack{\justify{ We acknowledge Harsh Gupta, Ritik Jain, Sulagna Saha, and Subhash Chaturvedi for their valuable comments on the preparation of this manuscript. 
The work done in this paper is supported by the grant USISTEF/QT/165/2023 received from the U.S.-India Science and Technology Endowment Fund (USISTEF).
A.R. acknowledges funding support from the Department of Science and Technology (DST), Government of India having grant number DST/QTC/NQM/QComm/2024/2 under the National Quantum Mission (NQM).}}



\data{No new data were created or analysed in this study.}


\bibliographystyle{iopart-num}
\bibliography{references}

\end{document}